\documentclass[hess, manuscript]{copernicus}
\usepackage{comment}
\usepackage{hyperref}
\usepackage{subcaption}

\newcommand{\modif}[1]{\textcolor{red}{#1}}
\newcommand\wdmax{\textcolor{blue}{\mbox{}}}

\newcommand\Qunit{$\mathrm{m}^{3}.\mathrm{s}^{-1}$}

\begin{document}
\nolinenumbers


\title{Towards Digital Twin in Flood Forecasting with Data Assimilation Satellite Earth Observations - A Proof-of-Concept in the Alzette Catchment}



\Author[1][thanh-huy.nguyen@list.lu]{Thanh Huy}{Nguyen} 
\Author[1]{Sukriti}{Bhattacharya}
\Author[1]{Jefferson S.}{Wong}
\Author[1]{Yoanne}{Didry}
\Author[1]{Duc Long}{Phan}
\Author[1]{Thomas}{Tamisier}
\Author[1]{Patrick}{Matgen}

\affil[1]{Luxembourg Institute of Science and Technology, L-4362 Esch-sur-Alzette, Luxembourg}





\runningtitle{TEXT}

\runningauthor{TEXT}

\received{}
\pubdiscuss{} 
\revised{}
\accepted{}
\published{}


\firstpage{1}

\maketitle

\begin{abstract}
Floods pose significant risks to human lives, infrastructure, and the environment. Timely and accurate flood forecasting plays a pivotal role in mitigating these risks. 
This study presents a proof-of-concept for a Digital Twin framework aimed at improving flood forecasting in the Alzette Catchment, Luxembourg. The approach integrates satellite-based Earth observations, specifically Sentinel-1 flood probability maps, into a particle filter-based data assimilation (DA) process to enhance flood predictions. 
By combining the GloFAS global flood monitoring and GloFAS streamflow forecasts products with DA using a high-resolution LISFLOOD-FP hydrodynamic model, the Digital Twin can provide daily flood forecasts for up to 30 days with reduced prediction uncertainties.
Using the 2021 flood event as a case study, we evaluate the performance of the Digital Twin in assimilating EO data to refine hydraulic model simulations and issue accurate forecasts.
While some limitations, such as uncertainties in GloFAS discharge forecasts, remain large, the approach successfully improves forecast accuracy compared to open-loop simulations. Future developments will focus on constructing more adaptively the hazard catalog, and reducing inherent uncertainties related to GloFAS streamflow forecasts and Sentinel-1 flood maps, to further enhance predictive capability. The framework demonstrates potential for advancing real-time flood forecasting and strengthening flood resilience.
\end{abstract}


\section{INTRODUCTION}
The growing frequency and intensity of extreme water-related events, exacerbated by climate changes \citep{IPCC2021,IPCC2021SPM,RN159}, call for advanced decision-support systems that can accurately predict and monitor environmental disasters and manage water resources efficiently \citep{brocca2024digital}.
Advancements in Earth Observation (EO), coupled with the swift progress in satellite data analysis and access to distributed computing and storage, open up exciting possibilities for the development of Digital Twins of the Earth systems (DTE). These Digital Twins hold the potential to transform disaster preparedness, allowing us to foresee extreme events and assess the effectiveness of various policy measures and mitigation strategies \citep{HOFFMANN2023100394}.  

A digital twin is a virtual representation of both physical systems (e.g. water, air, traffic, etc.) and assets (e.g. buildings, infrastructures, resources.) allowing simulations, tests, and predictions of planned actions almost in real-time \citep{rasheed2020digital}. 
Digital Twins are being developed using innovative Earth system models, data sources, and technologies, with the primary goal of enabling a wide range of users to explore the effects of climate change on different components of the Earth system and evaluate adaptation and mitigation strategies \citep{henriksen2022new,brocca2024digital,10.3389/fsci.2024.1383659}. 
Several structural institutional initiatives supporting this  exist in Europe, such as Destination Earth (DestinE) led by the European Commission together with ESA, EUMETSAT and ECMWF \citep{HOFFMANN2023100394}, or ESA Digital Twin Earth have funded by a large number of ESA Member States, including Luxembourg. 
These initiatives aim to create a digital model of the Earth to monitor the impacts of natural disasters and human activities, as well as anticipating extreme events, and provide policy to adapt climate-related challenges \citep{nativi2021digital}.
In Luxembourg, the national initiative by Thales Alenia Space contributed to this goal by integrating a new regionally developed component to the flood prediction DTE, which led to this research work. 
Within this framework, we propose here a specialized Digital Twin dedicated to flood disasters, demonstrated via a Proof of Concept (PoC) applied to a selected region. The pilot study for this research work focuses on the summer 2021 flood event over the Alzette Catchment, Luxembourg. Its primary goal is to enhance flood resilience by introducing an innovative inundation forecasting service that provides early warnings and enhances preparedness.



Luxembourg has always been sensitive to floods and the trend is increasing \citep{ciabatti2018floods,ludwig2023multi}.
Luxembourg has long been prone to flooding, and this risk is increasing over time \citep{ciabatti2018floods, ludwig2023multi}. 
Additionally, snow melt-induced flooding primarily affects the northern regions and the Moselle area. 
Historically, the country has experienced numerous winter floods, including major events in 1983 along the Moselle River and in 1993, 1995, 2003, and 2011 in the Sauer basin. 
Although floods are common, the scale of the damage they cause has been rising significantly. For instance, severe rainfall triggered floods in southern Luxembourg in May–June 2016 and in the northeast in July 2016, resulting in material damages worth hundreds of thousands of euros. 
This paper presents the methodology and results of the PoC focusing on the summer 2021 flood occurred in the Alzette catchment. We demonstrate the flood forecast capability thanks to data assimilation, and the impact of satellite EO data in improving the forecast accuracy.

\subsection{Digital Twin}
DTE has been recognized as a cutting-edge concept in Earth system modeling, as it presents a paradigm shift in global flood forecasting strategies. The integration of advanced numerical models, heterogeneous EO data, and artificial intelligence (AI) and high-performance computing (HPC) capabilities forms the backbone of DTE. By creating a digital replica of Earth's physical systems, including meteorological/atmospheric dynamics, hydrological processes, and land surface states and fluxes, DTE offers very-high accuracy simulation and prediction of global flood events. Furthermore, DTE facilitates scenario-based simulations, allowing stakeholders to explore \textit{what-if} scenarios, and assess the potential impacts of climate change, land-use changes, and/or infrastructure developments on flood patterns.

DTE models represent a groundbreaking innovation, providing digital replicas that enable real-time monitoring and simulation of Earth's processes with unprecedented spatiotemporal resolution. Such a concept of a DTE, particularly for modeling the water, energy, and carbon cycles, faced significant challenges even before the term was formally introduced. Early pioneering studies by \citep{wood2011hyperresolution,https://doi.org/10.1002/hyp.10391} recognized these difficulties, including issues such as process representation and parametrization, limited access to high-resolution EO data for model inputs and validation, substantial computational demands, and the complexity of incorporating human impacts on the water cycle. These foundational efforts laid the groundwork for the development of DTE models in hydrology. Indeed, \citep{brocca2024digital} made substantial advancements in creating the first high-resolution EO-based products for key hydrological variables, including soil moisture, precipitation, evaporation, and river discharge, etc. across large regions. Their DTE focused on representing terrestrial hydrologic processes and integrate observations of water fluxes and states \citep{10.3389/fsci.2024.1383659}.

\subsection{Data assimilation in flood studies}
Numerical hydrodynamic models used in flood forecasting are inherently imperfect due to uncertainties in both the model structure and its inputs, such as friction coefficients and boundary conditions (BCs). These uncertainties propagate into the model's outputs, including water levels and discharge, leading to potential inaccuracies in predictions. Such uncertainties can hinder optimal decision-making during emergency situations \citep{hostache2010assimilation,pappenberger2007fuzzy}.
A well-established approach to reducing these uncertainties is the periodic adjustment of models through the assimilation of available observational data \citep{asch2016data,Moradkhani2018}. Over time, advances in data assimilation (DA) have significantly enhanced the accuracy of flood simulations and forecasts \citep{Madsen2005, Neal2007, Neal2009, leedal2010visualization}. Water levels and discharges time-series measured at gauge stations installed along rivers are commonly used for hydraulic model calibration and validation.
DA methods, including variational and ensemble-based approaches like the Ensemble Kalman Filter (EnKF) \citep{evensen2003ensemble}, integrate these time-series observations with numerical models. By doing so, they correct the hydraulic state variables and reduce uncertainties in model parameters, such as channel and floodplain roughness, as well as inflow discharge \citep{Neal2007}, thereby improving the overall forecast reliability.

The benefits of assimilating in-situ data within floodplains and the influence of the layout of the observation network have been highlighted by \citep{Wesemael2018, Ziliani2019, munoz2022accounting}.
However, due to the high installation and maintenance costs, in-situ gauge stations providing water level data (typically installed by rivers) are limited to only a few locations within a catchment \citep{mason2012automatic}. This sparse spatial distribution poses a challenge for the accuracy of numerical models in both simulation and forecasting, especially within the floodplain.
This limitation can be mitigated by incorporating additional data sources, such as remote sensing (RS) flood extent maps, which, despite having a low revisit frequency, provide a 2D representation of flow dynamics. This broader coverage helps to improve model precision by supplementing the limited in-situ measurements.
\citep{Jafarzadegan2019} demonstrates how densifying the in-situ observing network improves the performance of the DA method built upon a LISFLOOD-FP model with a dual state-parameter EnKF strategy that account for correlations between point source observation errors. 

A traditional DA approach involves the assimilation of water surface elevation (WSE) data, which is a key diagnostic hydrometric variable in flood models. WSE data can be from in-situ gauge measurements, altimetry satellites, or derived from even RS images. In the case of RS, flood edge locations are combined with digital elevation models (DEM) to estimate WSE \citep{grimaldi2016remote,garambois2020variational,Dasgupta2021review}.
\citet{andreadis2007prospects} performed an EnKF framework that assimilates synthetical altimetry data with a LISFLOOD-FP hydrodynamic model to improve the estimation of river discharge and water depth, reducing errors by up to 50\% compared to Open Loop simulations without assimilation.
Satellite synthetic aperture radar (SAR) data is highly beneficial for flood studies due to its ability to provide all-weather, day-and-night global coverage of continental water bodies and flooded areas, typically characterized by low backscatter (BS) values caused by the specular reflection of radar pulses \citep{martinis2015flood}. Several studies have explored the assimilation of RS-derived WSE, as summarized in \citep[Table 1]{revilla2016integrating}. A commonly used approach, such as that presented by \citet{giustarini2011assimilating}, involves identifying flood edges on SAR images and integrating them with high-resolution DEM to estimate WSE over the floodplain. This WSE is then compared to or assimilated into the WSE produced by 1D or 2D hydrodynamic models to iteratively update the model's state and parameters. The assimilation of RS-derived WSE is advantageous as it directly uses a key diagnostic variable of the model \citep{giustarini2011assimilating,annis2021simultaneous}. However, its effectiveness may be limited by the vertical accuracy of the topographic data, even when the planimetric resolution is high, as highlighted in various studies \citep{giustarini2011assimilating,matgen2010towards,hostache2009water,garcia2015satellite}.
 
Retrieving WSE from flood extents can be bypassed by directly assimilating flood probability maps or flood extent maps derived from SAR images. Various methods have demonstrated the assimilation of surface water extents in both large-scale hydrology and catchment-scale hydrodynamics. \citet{lai2014variational} introduced a variational DA scheme built on a 2D Shallow Water model for assimilating flood extent observations obtained from MODIS data to correct roughness parameters across the floodplain. The assimilation of flood extent data has proven effective in improving flood modeling in floodplains or areas with gently sloping, slowly-varying bed profiles.
In \citep{revilla2016integrating}, daily surface water extents from the so-called Global Flood Detection System are assimilated with an EnKF that relies on the random perturbations of the precipitation---input to the distributed hydrologic rainfall-runoff LISFLOOD model \citep{van2010lisflood}---with a focus on Africa and South America catchments. 
It has been demonstrated that assimilating RS-derived surface water extents significantly enhances flood peak forecasting in terms of both timing and volume, particularly in slow-moving, ungauged catchments. While these research works \citep{revilla2016integrating,lai2014variational} rely on the expression of flood maps as a function of the model state, others propose a more direct use of SAR observations.

\citet{hostache2018near} presents the assimilation of ENVISAT ASAR-derived flood probability maps using a Particle Filter (PF) approach with a Sequential Importance Sampling (SIS) \citep{plaza2012importance} into a coupled hydrologic-hydraulic SUPERFLEX-LISFLOOD-FP model. 
The PF frameworks used in \citep{matgen2010towards,giustarini2011assimilating,hostache2018near,Dasgupta2020,DiMauro2021} advantageously allows relaxing the assumption that observation errors are Gaussian, and allows propagating a non-Gaussian distribution through non-linear hydrologic and hydrodynamic models  \citep{moradkhani2008hydrologic}. This makes it better suited for a DA of  flood maps than the more widely used EnKF \citep{garcia2015satellite,revilla2016integrating,Neal2007} or variational approaches \citep{lai2014variational}. 
As explained in \citep{giustarini2016probabilistic}, a probabilistic flood map indicates the likelihood that an observed BS value corresponds to a flood pixel, based on the assumption that the prior probabilities of being flooded or non-flooded follow two Gaussian probability density functions (PDF).
\citet{dimauro2022tempered} proposed a tempered PF to resolve the degeneracy and sample impoverishment problems of a classical PF \citep{kong1994sequential}, allowing to extend the PF benefits over time. 
\citet{GARCIAALEN2023129667} investigated the joint assimilation of SMAP soil moisture and streamflow data into a 2D shallow water model, demonstrating improved hydrological predictions through a PF-based data assimilation framework.

\section{METHOD}

This present work aims at generating state-of-the-art flood hazard maps by relying on various data inputs, especially relevant datasets such as the medium-range flood streamflow forecasts\footnote{\url{https://global-flood.emergency.copernicus.eu/technical-information/glofas-30day/}} \citep{harrigan2020daily,glofas_v3.1} and GFM\footnote{\url{https://global-flood.emergency.copernicus.eu/technical-information/glofas-gfm/}} (Global Flood Monitoring) from the Global Flood Awareness System (GloFAS). 
The GloFAS medium-range flood streamflow forecast (called henceforth GloFAS forecasts for the sake of simplicity) provides ensemble real-time daily forecasts for the next 30 days. The ensemble includes one control forecast and 50 member forecasts.
On the other hand, GFM provides on a regular basis different flood-related products, especially Sentinel-1 SAR-derived flood extent maps provided by LIST-developed HASARD\textsuperscript{\textcopyright} mapping algorithms \citep{matgen2010towards,giustarini2016probabilistic,chini2017hierarchical}.

\begin{figure}[h]
    \centering
    \includegraphics[width=0.85\linewidth]{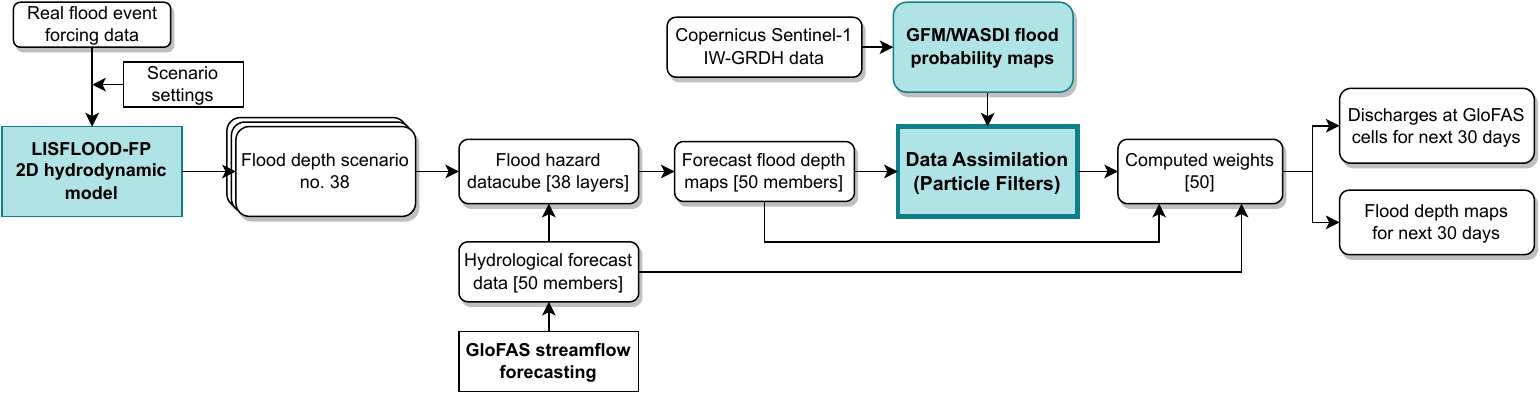}
    \caption{Diagram of the proposed method.}
    \label{fig:diagram}
\end{figure}

Flood forecasting systems often comprise coupled rainfall-runoff model and hydraulic model \citep{hostache2018near}. 
Being shown essential for effective flood management \citep{grimaldi2016remote}, this cascade setup is relevant for obtaining precise predictions of flood extent and water depth across large floodplains. In contrast, simplified channel flow routing models typically struggle to provide accurate predictions of floodplain inundation extents \citep{montanari2009calibration,schumann2013first}.
Figure~\ref{fig:diagram} illustrates the diagram of the proposed flood forecasting system. In this research work, the flood forecasting strategy involves the local grid-based hydraulic model LISFLOOD-FP combined with inflow discharges provided by GloFAS streamflow forecasts.  
However, the GloFAS forecast discharge ensemble is not directly used as forcing data for the LISFLOOD-FP model. Instead, the hydrodynamic LISFLOOD-FP model is employed to compute beforehand a series of flood depth maps based on various scenario settings that account for a wide range of meteorological and hydrological conditions. The GloFAS forecast discharge is then used as a reference to identify the corresponding pre-computed flood depth maps, rather than serving as direct forcing data for real-time simulation. The DA process is implemented using a PF approach to assimilate flood extent maps that are derived from Sentinel-1 images using GFM. 


High-resolution numerical simulations by LISFLOOD-FP (described below in subsection~\ref{ssec:lisflood}) pre-compute maximum water depth maps for various river inflow discharge scenarios, forming a flood hazard datacube with 38 layers (described below in subsection~\ref{ssec:scenario}). Each layer corresponds to a specific discharge scenario, and the number of layers is associated with the number of peak discharges used by the scenarios. Using the GloFAS probabilistic discharge forecast ensemble (described below in subsection \ref{ssec:glofas}), the forecast streamflow discharge is matched with the datacube at a daily time step to extract corresponding flood depth maps at 5-meter resolution, generating an ensemble of flood inundation maps based on the forecasts. Each shall be considered a \textit{particle}.
When available, Sentinel-1-derived flood extent maps, processed by GFM/WASDI (described in subsection \ref{ssec:gfm}), are used to compare with the simulated flood extent maps from the ensemble to update the weight of the particles according to their agreement with the Sentinel-1 observed flood extent maps.

Incorporating satellite-based flood observations is a key feature of this strategy. Sentinel-1 flood observations, processed by flood extent mapping methods, provide NRT and reliable evidence of flood dynamics. These satellite-based observations are particularly valuable in regions where ground-based monitoring data is sparse or even unavailable. The DA compares the flood extent maps from either GFM (called PF1) or from WASDI-based urban flood mapping \citep{chini2019sentinel} (called PF2), with the pre-computed flood hazard datacube generated by LISFLOOD-FP while resampling them into the same spatial resolution. Through this comparison, the PF assigns higher weights to simulations that align more closely with the observed flood extents, further refining the predicted flood depth distribution.




By integrating the GFM and GloFAS products through DA, built on top of a local LISFLOOD-FP hydrodynamic model with high spatial resolution outputs, the Digital Twin is capable of medium-range daily inundation forecasting up to 30 days, while reducing predictive uncertainties. The DA strategy is flexible and accommodates various global- and local-scale models and resolutions. The PF implementation involves the weighting scheme as proposed by \citet{dimauro2022tempered} to deal with degeneracy and impoverishment problems.
It allows weighted combinations of pre-computed flood depth maps from LISFLOOD-FP according to its alignment with flood extent maps observed by Sentinel-1.



\subsection{Study Area, Data, and Hydraulic Model}\label{ssec:lisflood}



\begin{figure}[h]
    \centering
    \begin{minipage}{0.48\linewidth}
    \centering
    \begin{subfigure}{\linewidth}
    \centering
    \includegraphics[width=0.8\linewidth]{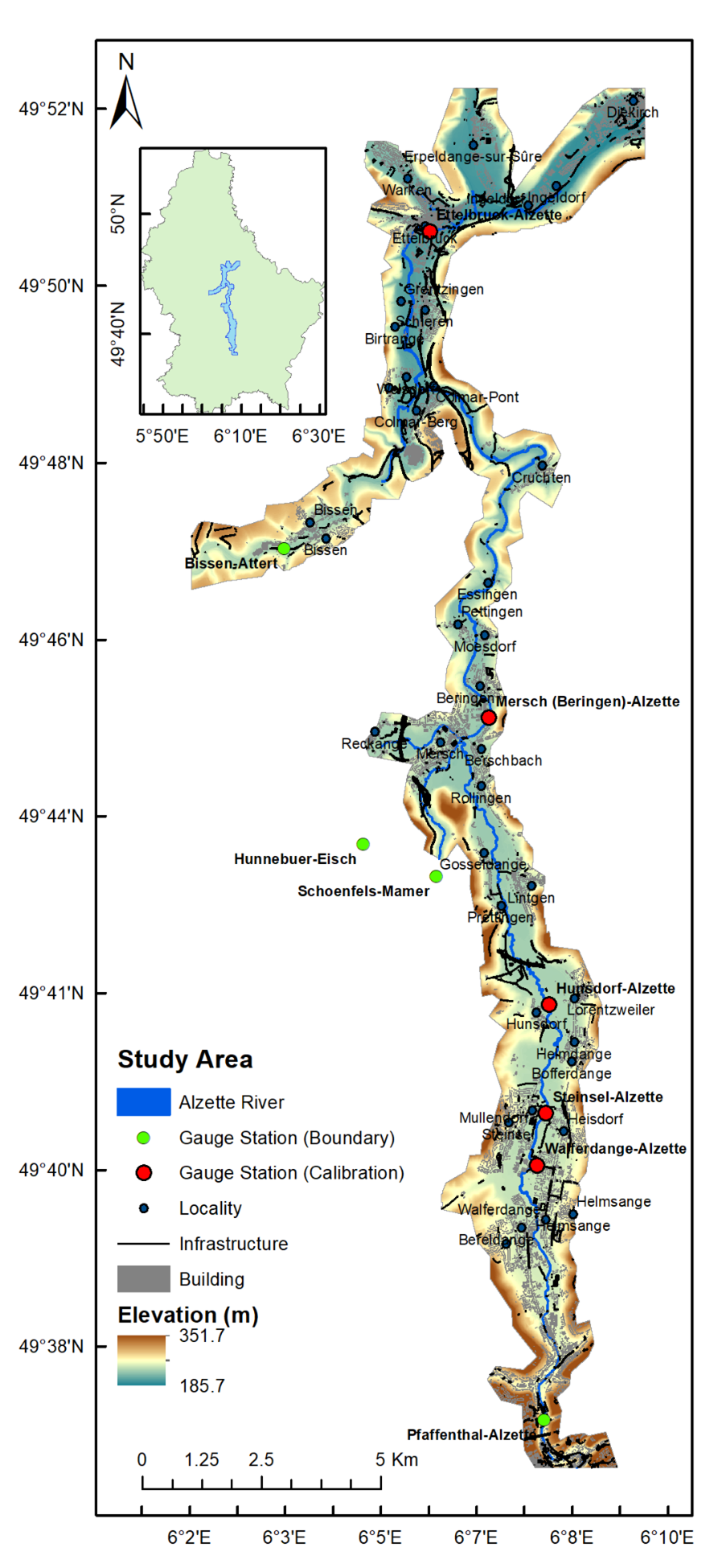}
    \caption{Study area.}
    \label{fig:model}
    \end{subfigure}
    \end{minipage}
    \hfill
    \begin{minipage}{0.48\linewidth}
    \begin{subfigure}{\linewidth}
    \includegraphics[trim=0 0.25cm 0 0, clip, width=0.95\linewidth]{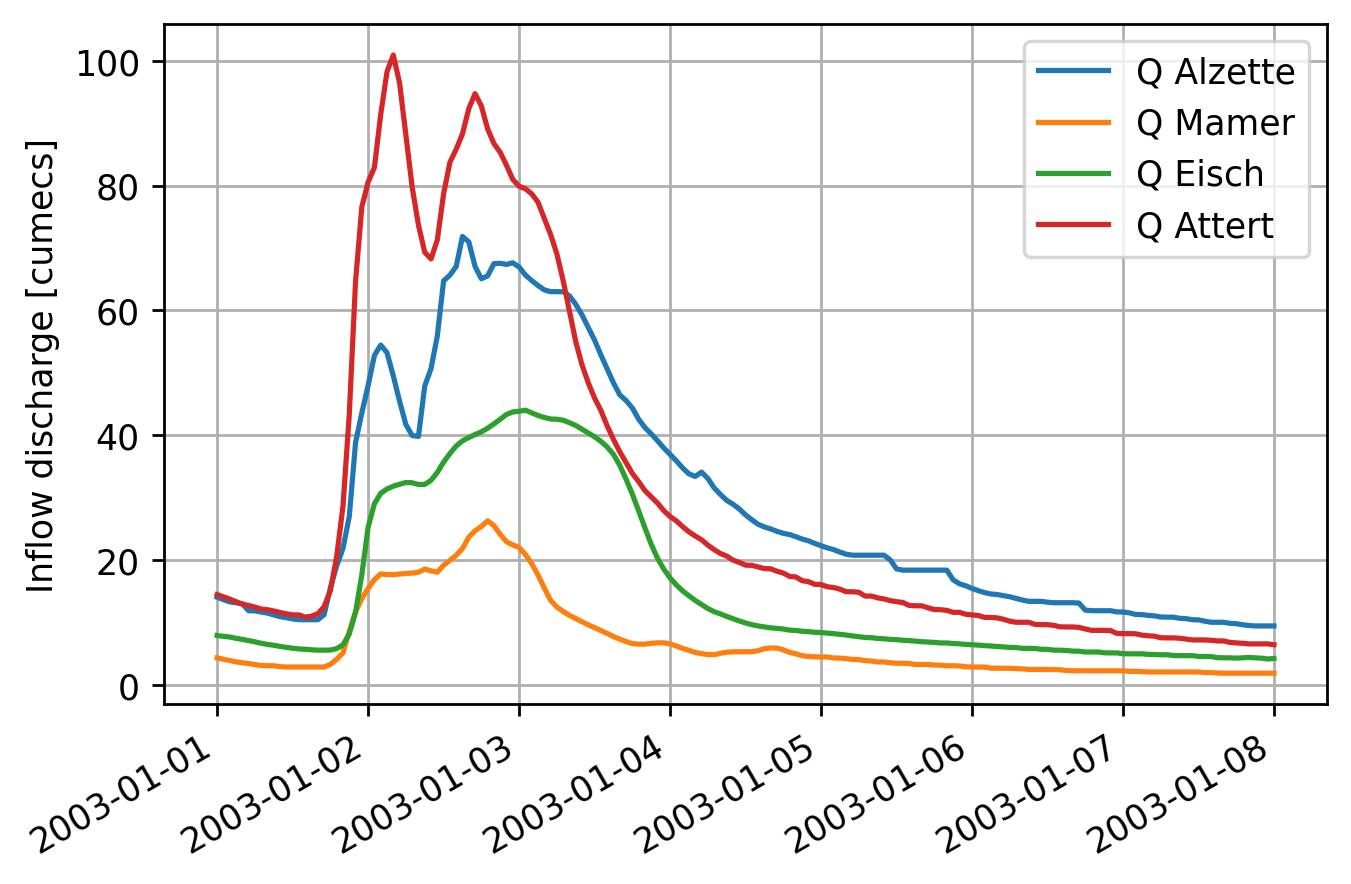}
    \caption{2003 flood event (used for calibration).}
    \label{fig:Q2003_Alzette}
    \end{subfigure}

    \vspace{0.5cm}
    
    \begin{subfigure}{\linewidth}
    \includegraphics[trim=0 0.25cm 0 0, clip, width=0.95\linewidth]{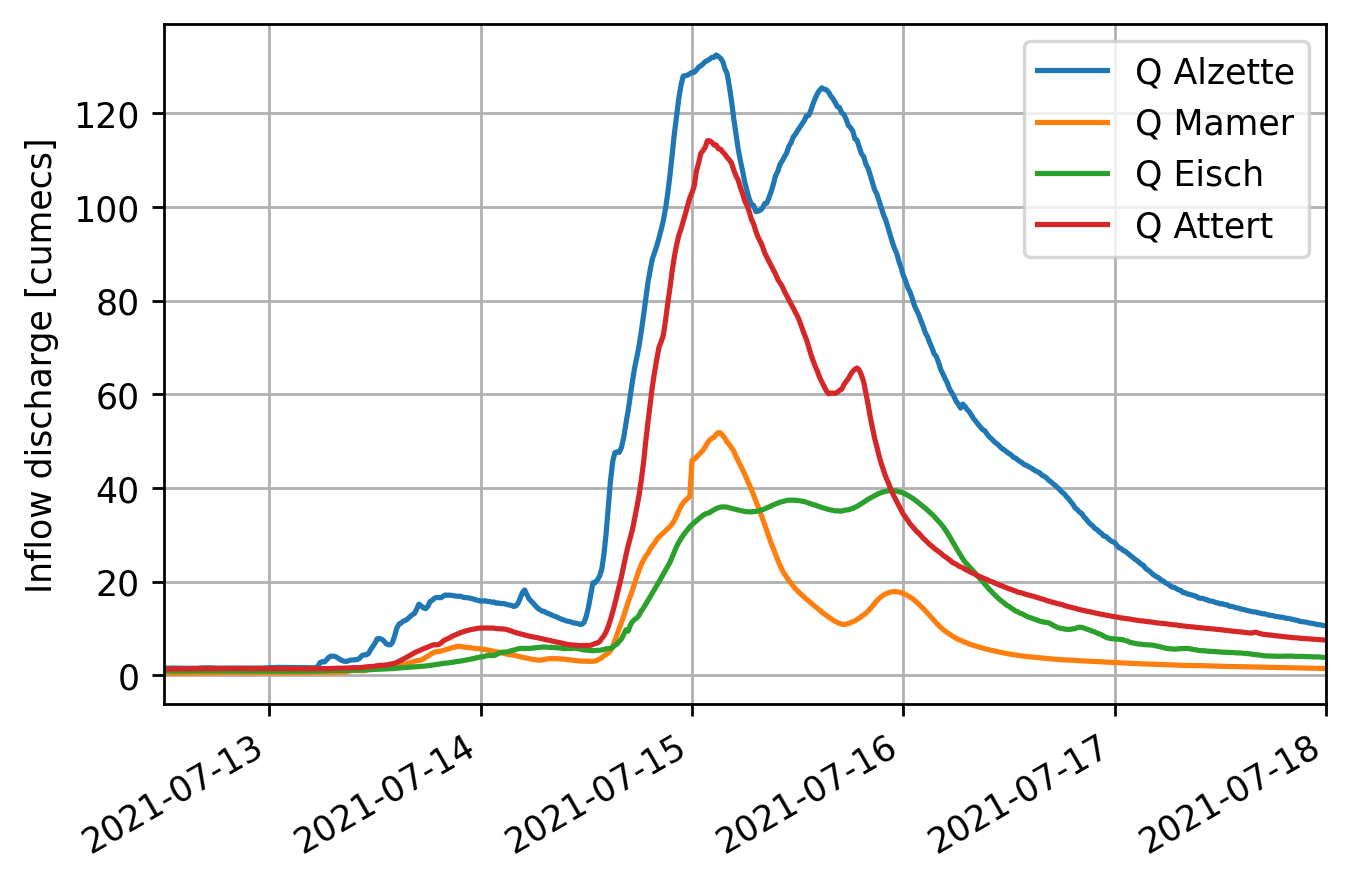}
    \caption{2021 flood event.}
    \label{fig:Q2021_Alzette}
    \end{subfigure}
    \end{minipage}
    \caption{(a) The study area on the Alzette River  (Luxembourg). The study reach is located downstream of Luxembourg City between the gauge stations at Pfaffenthal and Ettelbruck. Green solid dots represent gauge stations providing upstream boundary conditions, whereas red solid dots represent those used for the calibration and validation. Inflow discharge for (b) 2003 and (c) 2021 flood events at Pfaffenthal (Alzette), Schoenfels (Mamer), Hunnebuer (Eisch), and Bissen (Attert).}

\end{figure}



\subsubsection{Study Area and Gauging Station Network}


The study area on the Alzette River (Luxembourg) is situated downstream of Luxembourg City, between the gauging stations at Pfaffenthal (upstream) and Ettelbruck (downstream), covering a stretch of approximately 28 km with an average slope of 0.001 m/m. This section is also fed by three tributaries: the Mamer, Eisch, and Attert Rivers. Figure~\ref{fig:model} illustrates the model domain and river gauge network, along with the catchment topography. Green dots in Figure~\ref{fig:model} mark the model's upstream boundary condition (BC) inputs, located at Pfaffenthal (Alzette River), Schoenfels (Mamer River), Hunnebour (Eisch River), and Bissen (Attert River). The information on these gauge stations is summarized by Table~\ref{tab:BC_stations}. On the other hand, red dots in Figure~\ref{fig:model} represent additional gauging stations within the model domain used for model calibration and validation. They are summarized by Table~\ref{tab:Cal_stations}, providing discharge and water level time-series at a 15-minute time step throughout the studied flood events.
In the context of flooding, Table~\ref{tab:return_period} in Appendix \ref{app2} summarizes the statistical discharge and water level values at these stations according to the return period.

\begin{table}[h]
    \centering
    \caption{Gauging stations for upstream BCs.}
    \label{tab:BC_stations}
    \scalebox{0.9}{
    \begin{tabular}{cccccccccc}
        \hline
        & Water- & Basin & Gauge & River & X & Y & Operating & Early alert & Alert \\
        Name (Operator) & course & size [km$^2$] & altitude [m NN.] & kilometre [km] & (LUREF) & (LUREF) & since & threshold [m] & threshold [m] \\\hline
        Pfaffenthal (AGE) &  Alzette & 360.5 & 235.25 & 35.04 & 77409 & 76226 & 01.10.1996 & 2.20 & 2.70 \\
        Schoenfels (AGE) & Mamer & 83.6 & 222.56 & 3.79 & 75145 & 87624 & 01.09.1996 & - & - \\
        Hunnebour (AGE) & Eisch & 164.2 & 223.48 & 5.33 & 73612 & 88301 & 01.08.1996 & 2.35 & 2.80 \\
        Bissen (AGE) & Attert & 291.5 & 218.56 & 5.6 & 71955 & 94495 & 01.12.1996 & 2.50 & 3.0\\\hline
    \end{tabular}}
\end{table}

\begin{table}[h]
    \centering
    \caption{Gauging stations on the Alzette for calibration/validation.}
    \label{tab:Cal_stations}
    \scalebox{0.9}{
    \begin{tabular}{cccccccccc}
        \hline
        & Water- & Basin & Gauge & River & X & Y & Operating & Early alert & Alert \\
        Name (Operator) & course & size [km$^2$] & altitude [m NN.] & kilometre [km] & (LUREF) & (LUREF) & since & threshold [m] & threshold [m] \\\hline
        Walferdange (AGE) &  Alzette & 405.1 & 225.32 & 28.69 & 77256 & 81572 & 01.11.2002 & - & - \\
        Steinsel (AGE) &  Alzette & 406.9 & 222.26 & 27.43 & 77432 & 82659 & 01.11.1996 & 3.0 & 3.50 \\
        Hunsdorf (LIST) &  Alzette & 410.1 & 220.57 & 24.69 & 77510 & 84935 & 29.06.2001 & - & - \\
        Mersch (AGE) &  Alzette & 707.0 & 212.35 & 16.48 & 76243 & 90955 & 01.08.1996 & 3.50 & 4.0 \\
        Ettelbruck (AGE) &  Alzette & 1091.9 & 193.99 & 1.14 & 74998 & 101159 & 01.09.1996 & 1.80 & 2.30 \\\hline
    \end{tabular}}
\end{table}





\subsubsection{LISFLOOD-FP Hydraulic Model}\label{sssec:lisflood}
LISFLOOD-FP \citep{bates2000simple} is a raster-based 2D hydraulic model that simulates the propagation of flood waves along channels and across floodplains using a storage cell approach. The model solves the Shallow Water Equations (SWEs) at low computational cost by neglecting the convective acceleration term using an explicit finite difference scheme \citep{bates2010simple} and the model time step is determined by the Courant-Friedrichs-Lewy condition to ensure numerical stability \citep{de2012improving}. A sub-grid river channel representation is introduced to allow the characterization of channel width independent of the nominal model resolution \citep{neal2012subgrid}.

Setting up a LISFLOOD-FP model requires a DEM for floodplain topography, channel geometry (river width, depth, and shape), BCs, and Manning's  roughness coefficients $n$ for both the channel and floodplain. The BCs consist of upstream discharge time series at all inflow points to the model domain and downstream water surface elevation time series at all outflow points
Three additional input data (channel width, channel depth, and bank elevation) are required for the sub-grid channel scheme.
A sub-grid formulation of LISFLOOD-FP was set up in this study, with the channel of the Alzette River represented as 1D sub-grid scale features and the floodplain using the inertial formulation of the shallow water equations \citep{bates2010simple}. The floodplain topography and three tributaries (Mamer, Eisch, and Attert) in the model were described using the 2019 LiDAR DEM\footnote{\url{https://data.public.lu/fr/datasets/lidar-2019-modele-numerique-de-terrain-mnt/}}  rescaled to a 5-meter spatial resolution, balancing computational efficiency for multiple simulations required for the calibration process with the model ability to accurately capture floodplain inundation patterns.

The width of the 1D sub-grid channel was derived from 266 surveyed cross-section widths\footnote{\url{https://map.geoportail.lu/theme/main?version=3&zoom=9&lang=fr&layers=2494&opacities=1&rotation=0&X=644262&Y=6394482&time=&bgLayer=basemap_2015_global}}, provided by the AGE within the framework of the "Cartographie des profils en travers 2021" . The DEM elevation was used to represent the bank height elevation, defining the channel bankfull depth in combination with channel bed elevation \citep{bates2013lisflood}. The channel geometry was assumed to be rectangular, since similar water level simulation accuracy was observed between models with rectangular channels and calibrated channel roughness and those with non-rectangular shapes \citep{neal2015efficient}. Given this configuration, the channel depths in the model may differ from in-situ measurements, and thus were approximated during calibration using a empirical power law formulation \citep{neal2012subgrid}, given by $D=r_{ch} W^{p_{ch}}$ where $D$ is the channel depth, $W$ the channel width, and $r_{ch}$ and $p_{ch}$ are the channel depth parameters. It is acknowledged that the rectangular channel configuration is unlikely to significantly impact the simulation results \citep{neal2015efficient}. 

Regarding the BCs, hourly discharge data from four gauging stations on Alzette at Pfaffenthal, Mamer at Schoenfels, Eisch at Hunnebuer, and Attert at Bissen (Figure~\ref{fig:Q2003_Alzette} and
Figure~\ref{fig:Q2021_Alzette} for the 2003 and 2021 flood event, respectively) were used as the upstream inflow points. On the other hand, instead of utilizing the gauging station at Ettelbruck as a downstream stage-varying BC, a free downstream BC was imposed to account for the backwater effect. It should be noted that the 2003 flood event (Figure~\ref{fig:Q2003_Alzette}) is only used for the calibration of the LISFLOOD-FP model, based on in-situ data and reference flood hazard maps. On the other hand, the 2021 flood event (Figure~\ref{fig:Q2021_Alzette}) is the focus of the paper, demonstrating the merits of the assimilation of satellite EO data in the context of flood forecasting.

Together with the two channel depth parameters $r_{ch}$ and $p_{ch}$, the Manning's $n$ roughness coefficients for channel ($n_{ch}$) and floodplain ($n_{fp}$) were also considered in the calibration process. A priori parameter distributions were assumed to be uniform due to insufficient evidence for effective parameter distributions \citep{aronica2002assessing,hunter2005utility}. The parameter ranges were determined based on their physically feasible limits and values reported in the literature. If no values were available in literature, default values based on expert knowledge were used, with a variation of $\pm 50\%$ from the default or within ranges suggested by prior modeling experience. 
The Alzette River in this study area is characterized as a gravel bed, concrete banks, and a floodplain dominated by cultivated areas and brush. Therefore, to capture the full range of potential roughness values, channel roughness $n_{ch}$ values was set between 0.01 and 0.05, and floodplain roughness $n_{fp}$ between 0.03 and 0.15 \citep{te1959open}. The ranges for $r_{ch}$ and $p_{ch}$ were set between 0.01 and 0.15, and between 0.01 and 1.00, respectively. 
The distribution of these parameters are summarized in Table~\ref{tab:params}. 
Latin hypercube sampling \citep{helton2003latin} was employed to enhance the efficiency of Monte Carlo sampling, ensuring effective coverage of parameter interactions in multiple dimensions while generating minimally correlated parameter sets from a sparse sample size \citep{beven2001equifinality}. In total, 500 parameter sets were generated for this study.

\begin{table}[h]
    \centering
    \caption{Distribution for LISFLOOD-FP parameters.}
    \label{tab:params}
    \begin{tabular}{ccc}
        Parameter & Distribution \\
        \hline
        Channel depth $r_{ch}$ & $\mathcal{U}\left(0.01; 0.15\right)$  \\
        Channel depth $p_{ch}$ & $\mathcal{U}\left(0.01; 1.00\right)$  \\
        Channel roughness $n_{ch}$ & $\mathcal{U}\left(0.01;0.05\right)$  \\
        Floodplain roughness $n_{fp}$ & $\mathcal{U}\left(0.03; 0.15\right)$  \\
        \hline
    \end{tabular}
\end{table}

Two observed datasets were used for model assessment: 
1) stage data from the Walferdange and Steinsel stations, and discharge data from the Hunsdorf, Mersch, and Ettelbruck stations, and 
\modif{2) a flood inundation extent map extracted from the Sentinel-1 satellite imagery}. 
The model performance in reproducing the timing and magnitude of the January 2003 flood event was evaluated against the gauging stations using the Kling-Gupta efficiency (KGE) \citep{gupta2009decomposition}, defined in subsection \ref{sect:metrics}, while its ability to capture the flood inundation extent of the July 2021 flood event was assessed by comparing the flood extent map and calculating the Critical Success Index ($\mathrm{CSI}$). Based on $\mathrm{KGE}$ and $\mathrm{CSI}$ results, the parameter sets were ranked and normalized into generalized conditional probabilities \citep{beven2001equifinality}. The observed datasets were combined by evaluating on one dataset and using the other to update the weights for each parameter set through a recursive application of Bayes' equation \citep{beven2001equifinality}. The parameter set associated with the highest combined conditional probability values was selected as the best parameter set for generating the water depth and flood inundation extent maps.   

\subsection{Scenario generation}\label{ssec:scenario}

\begin{figure}[h]
    \centering
    \includegraphics[width=0.5\linewidth]{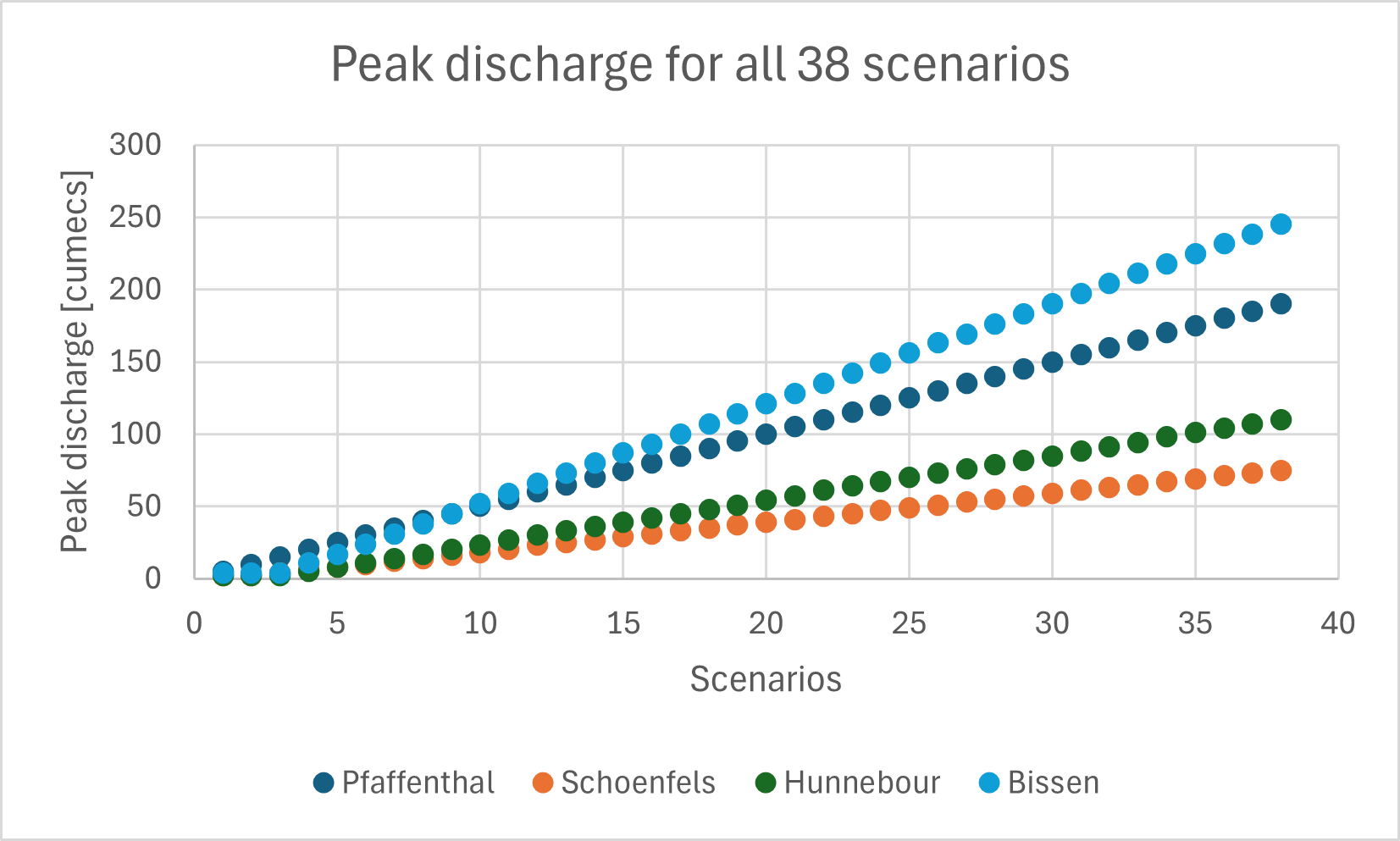}
    \caption{Peak discharge for all 38 scenarios for all four BCs.}
    \label{fig:Q_scenarios}
\end{figure}

\begin{figure}[h]
    \centering
    \centering
    \begin{subfigure}{0.24\linewidth}
    \centering
    \includegraphics[trim=0 0 17cm 0,clip,width=\linewidth]{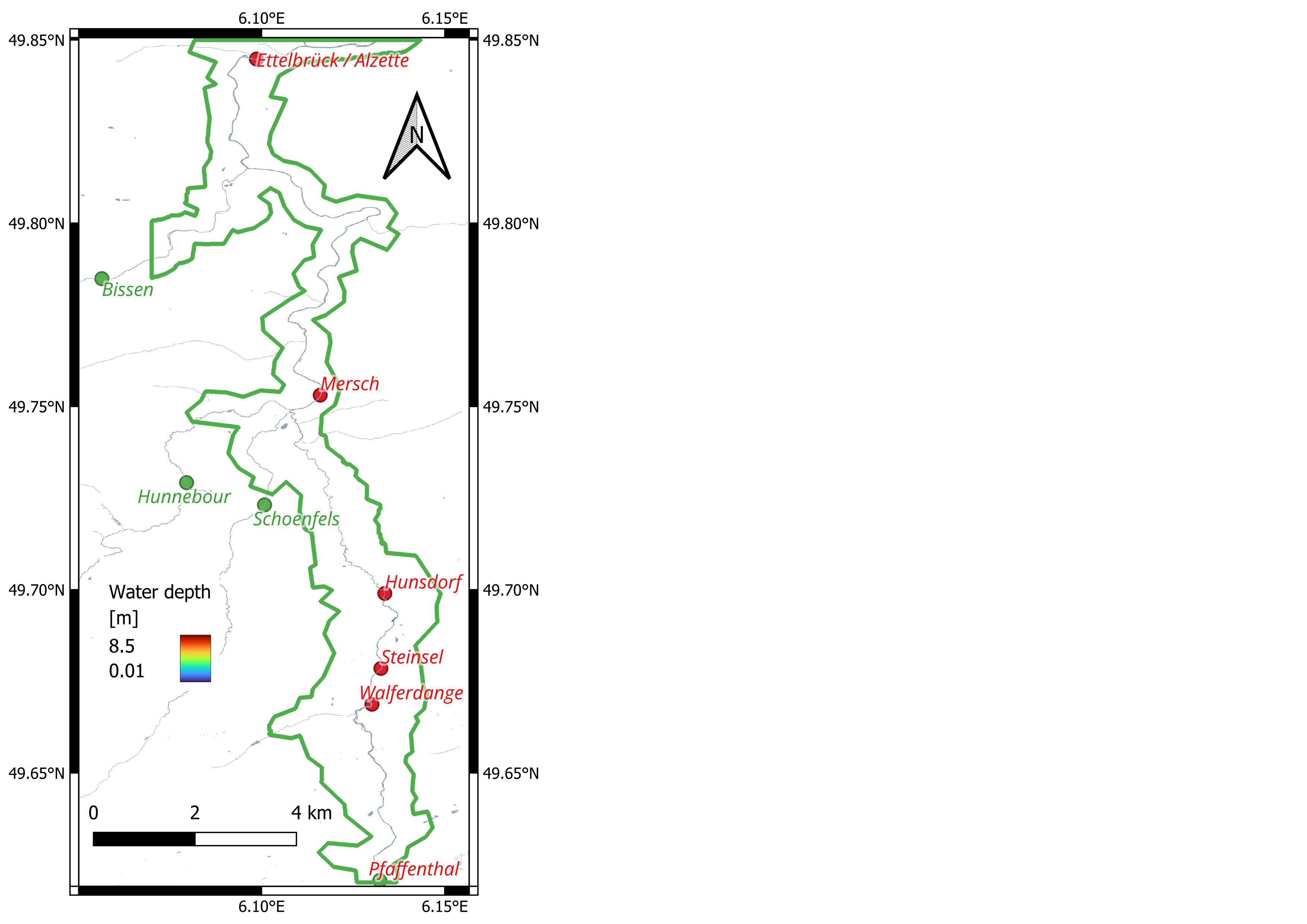}
    \caption{S1}
    \end{subfigure}
    \begin{subfigure}{0.24\linewidth}
    \centering
    \includegraphics[trim=0 0 17cm 0,clip,width=\linewidth]{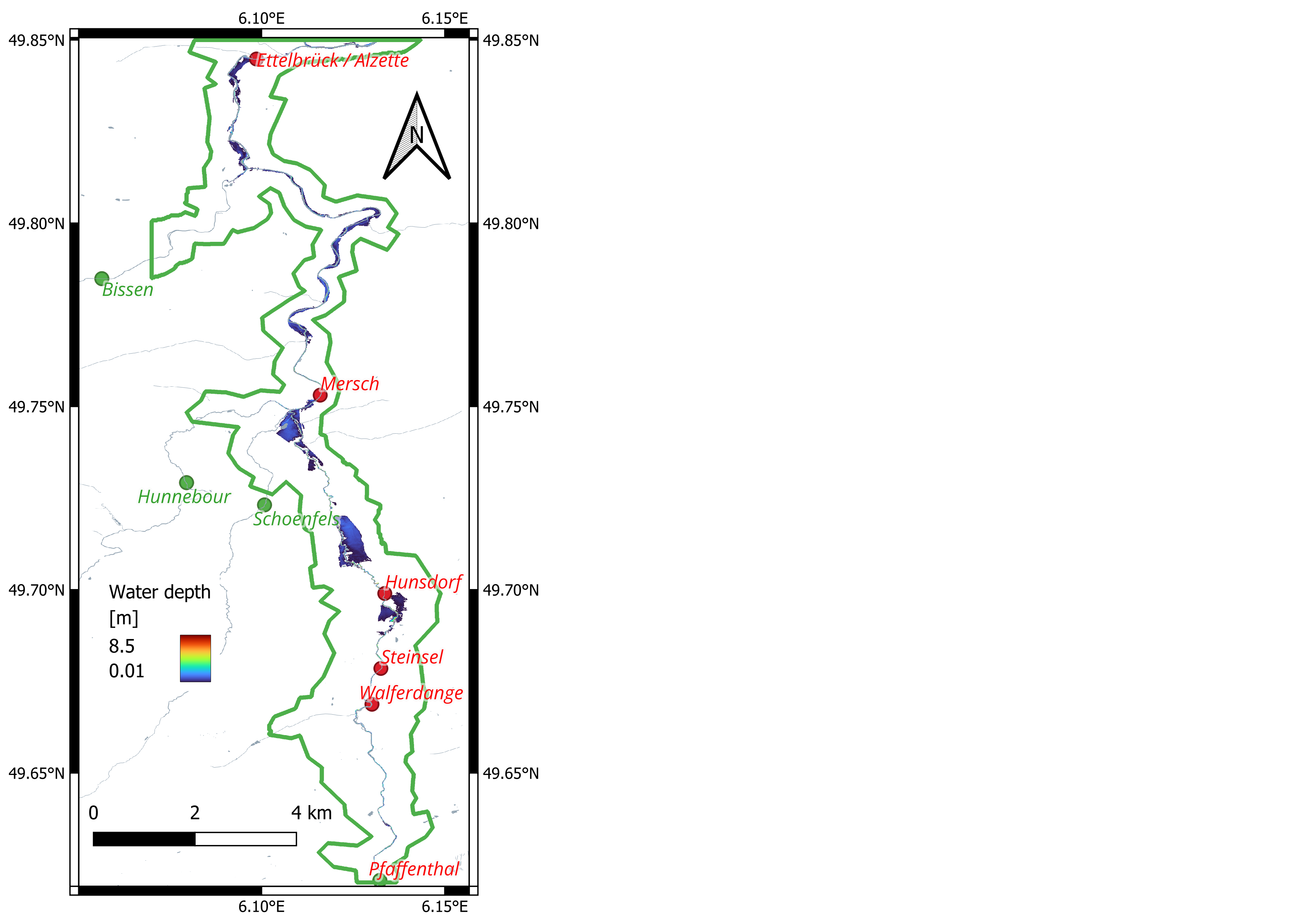}
    \caption{S14}
    \end{subfigure}
    \begin{subfigure}{0.24\linewidth}
    \centering
    \includegraphics[trim=0 0 17cm 0,clip,width=\linewidth]{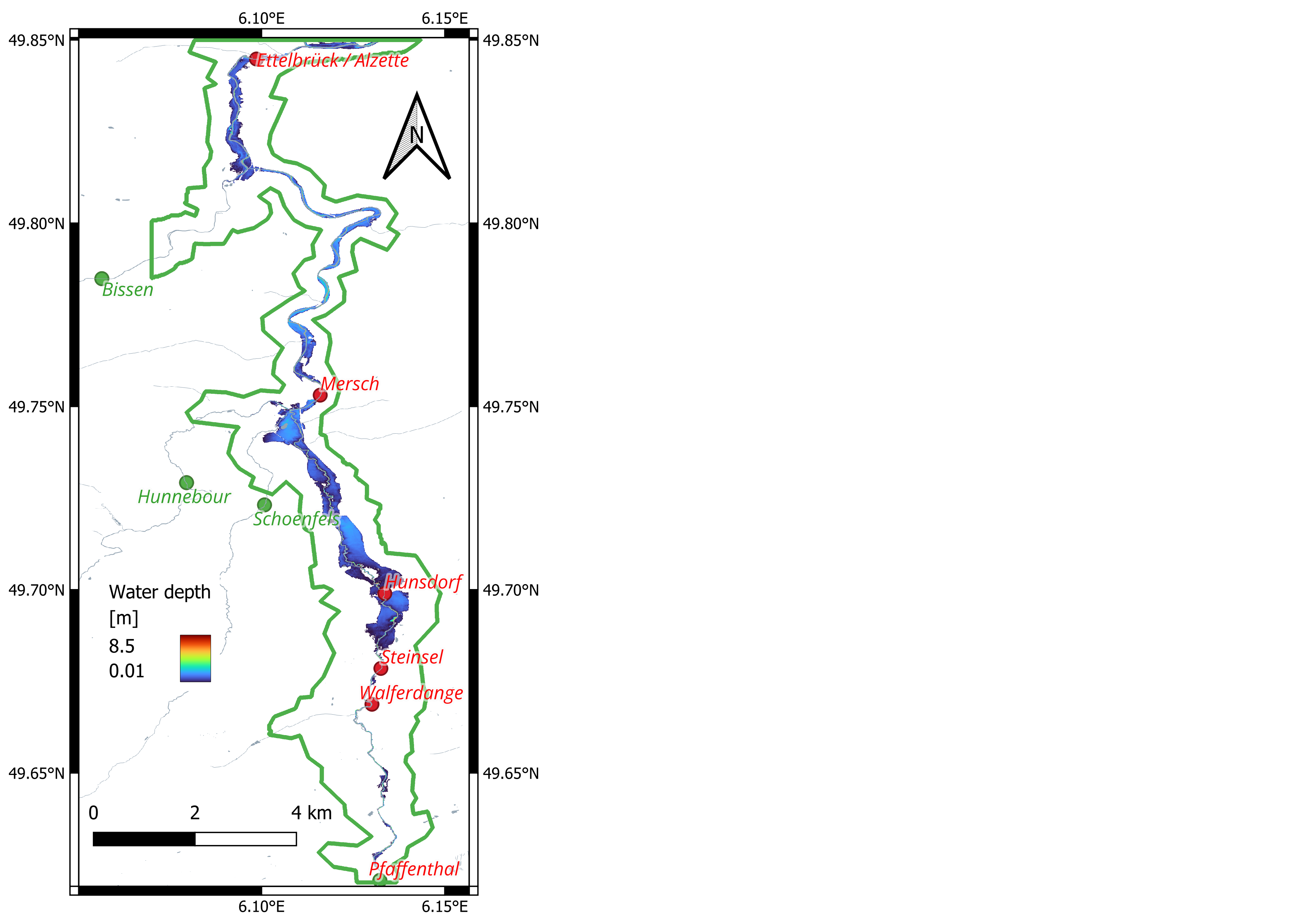}
    \caption{S27}
    \end{subfigure}
    \begin{subfigure}{0.24\linewidth}
    \centering
    \includegraphics[trim=0 0 17cm 0,clip,width=\linewidth]{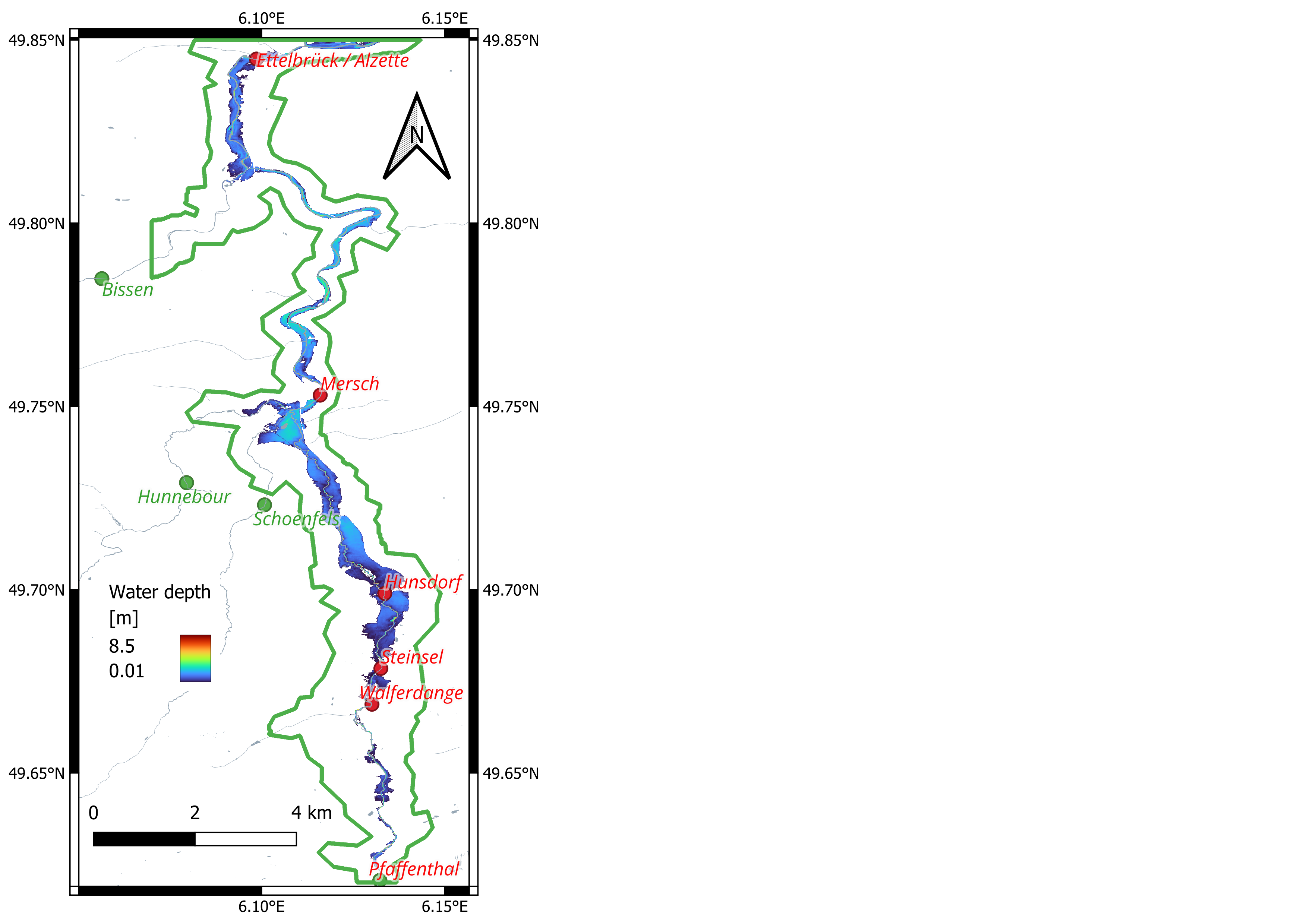}
    \caption{S38}
    \end{subfigure}
    \caption{Pre-computed 5-m resolution flood depth maps for different inflows using LISFLOOD-FP model. From left to right: four scenarios S1 (peak discharge at Pfaffenthal: 5~\Qunit), S14 (peak discharge at Pfaffenthal: 70~\Qunit), S27 (peak discharge at Pfaffenthal: 135~\Qunit) and S38 (peak discharge at Pfaffenthal: 190~\Qunit).}
    \label{fig:scenarios}
\end{figure}

A total of 38 scenarios were generated by scaling the maximum values of the upstream boundary condition at Pfaffenthal, represented by $Q_{max}$, with discharge values ranging from 5 to 190~\Qunit (with a step of 5~\Qunit). 
Each scenario reflects a different flow condition, allowing for a comprehensive analysis of hydrodynamic behavior across a wide range of potential discharge rates. The scenarios provide a detailed spectrum of flow responses for various hydraulic conditions, enabling accurate assessments of system performance under both low and high flow events. The four BCs (Table~\ref{tab:BC_stations}) were matched and rescaled based on their return periods (summarized by Table~\ref{tab:return_period} in Appendix~\ref{app2}).
In other words, each of the 38 scenarios corresponds to one $Q_{max}$ value at Pfaffenthal and thus one return period value. Such a return period value is then used to interpolate the corresponding $Q_{max}$ at the other three BCs.
Figure~\ref{fig:Q_scenarios} shows the peak discharge $Q_{max}$ values at all four BCs for all 38 scenarios. Such a setting is used to scale the 2021 flood event hydrograph into the potential flood scenarios. Running with the scaled hydrographs as forcing data, the LISFLOOD-FP model creates flood pre-computed water depth maps (illustrated by Figure~\ref{fig:scenarios}), constituting the flood hazard datacube.

\subsection{Hydrological rainfall-runoff model - GloFAS streamflow forecasts}\label{ssec:glofas}







GloFAS, within the CEMS, has been pre-operational since 2011 and fully operational since April 2018. 
Its hydrological modelling components include the land surface model of ECMWF IFS, H-TESSEL \citep{Balsamo2009}, and open-source LISFLOOD\footnote{\url{https://ec-jrc.github.io/lisflood/}} hydrological model \citep{van2010lisflood}.
To produce and predict global river discharge \citep{glofas_v3.1}, GloFAS uses medium- and extended-range meteorological forecasts from ECMWF-ENS as forcing data to the LISFLOOD, to provide publicly available ensemble real-time daily forecasts for the next 30 days for its global river network. More information can be found in Appendix~\ref{app:glofas}.

\begin{figure}[h]
    \centering
    \includegraphics[width=0.6\linewidth]{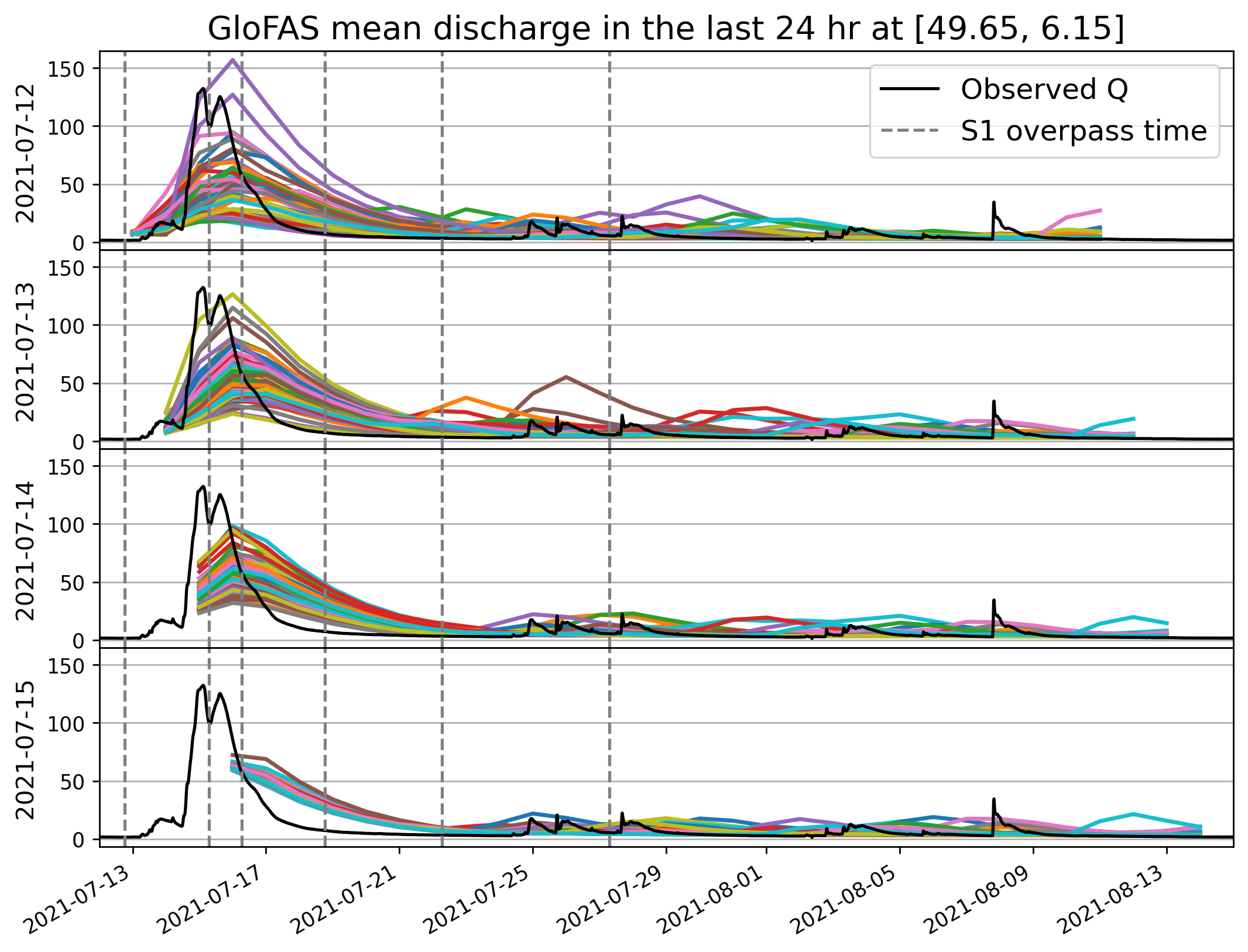}
    \caption{GloFAS streamflow forecasts issued on July 12 (first panel), July 13 (second panel), July 14 (third panel) and July 15 (fourth panel). Observed discharge (black lines) and 50 forecasted daily discharge (colored lines). The overpass times of Sentinel-1 SAR are indicated with vertical dashed lines.}
    \label{fig:Q_GloFAS}
\end{figure}

This gridded dataset offers a consistent representation of key hydrological variables across the global domain, namely river discharge, soil wetness index (root zone), snow water equivalent, runoff water equivalent (surface plus sub-surface). In this research work, we take advantage of the river discharge in the last 24 hours (i.e. average discharge over a 24-hour period) which is volume rate of water flow, including sediments, chemical and biological material, in the river channel averaged over a time step through a cross-section.  In this research work, we rely on the GloFAS streamflow forecast dataset (version 3.1) with a spatial resolution of $0.1^\circ \times 0.1^\circ$  \citep{glofas_v3.1}.
Figure~\ref{fig:Q_GloFAS} illustrates the daily GloFAS streamflow forecasts, generated with 50 ensemble members, starting from four issue dates: July 12, 2021 (first panel), July 13, 2021 (second panel), July 14, 2021 (third panel), and July 15, 2021 (last panel). It is important to note that each forecast begins for the day 1 after its issuance (e.g., the July 12 forecast corresponds to July 13 as day 1). It can be observed that forecast uncertainties increase as the flood peak approaches. It should also be noted that the July 15 forecasts (day 1 starting from July 16) in Figure~\ref{fig:Q_GloFAS} mark the recession of flood event and therefore they will not be included in the implementation of the proposed method.

\subsection{Flood extent mapping - GloFAS global flood monitoring (GFM)}\label{ssec:gfm}
Sentinel-1, a Copernicus constellation \citep{torres2012gmes} of two polar-orbiting satellites (up until December 2021\footnote{Sentinel-1B became defective in December 2021, and Sentinel-1C has been launched recently on December 2024.}), operates day and night performing C-band SAR imaging regardless of the weather conditions. Based on the systematic stream of Sentinel-1 data, near-real-time (NRT) GFM product can be provided for the Copernicus Emergency Management Service (CEMS) \citep{salamon2021new} within less than eight hours after image acquisition. The product is obtained by an ensemble approach combining the results of three independently developed flood mapping methods by \citep{chini2017hierarchical}, \citep{martinis2015fully}, and \citep{bauer2022satellite}.



Sentinel-1 data are collected in Interferometric Wide-swath mode with Ground Range Detected at High resolution (Sentinel-1 IW GRDH). The GRD products include focused SAR data that has been detected, multi-looked, and projected to ground range using an Earth ellipsoid model, with phase information discarded. This results in a product with approximately square pixels and spacing, offering reduced speckle but at the cost of lower spatial resolution. In this case, the raw backscatter amplitude is sampled with a pixel size of $10 \times 10$ meters. Only VV-polarization data are used here due to their superior sensitivity in distinguishing water from non-water surfaces.


The preprocessed Sentinel-1 output is formatted as Analysis-Ready-Data, which is then sent to the GFM flood detection engine. Here, three algorithms run concurrently to generate NRT flood mapping products. The GFM product relies on an ensemble approach that combines three well-established, independently developed flood detection algorithms from the Luxembourg Institute of Science and Technology (LIST) \citep{chini2017hierarchical}, the German Aerospace Center (DLR) \citep{martinis2015fully}, and Vienna University of Technology (TUW) \citep{bauer2022satellite}. The flood ensemble is calculated pixel by pixel using a majority voting system, where at least two of the algorithms must classify a pixel as flooded or non-flooded \citep{krullikowski2023estimating}.

A second method of flood extent mapping \citep{chini2019sentinel} has also been integrated in this work to allow a better mapping of floodwater in urban area, compared to GFM. 
It takes advantage of the Interferometric SAR coherence to detect the presence of floodwater in urbanized areas.
Such a method has been made available via WASDI platform\footnote{\url{https://www.wasdi.net/\#/edrift_auto_urban_flood/appDetails}}.

\subsection{Data Assimilation}
This section establishes the DA framework for integrating SAR-derived probabilistic flood maps within the flood forecasting model built around the hydraulic LISFLOOD-FP model.
Using the ensemble of GloFAS streamflow forecasts (comprising of 50 members) to match with the upstream BCs devised for each scenario, an ensemble of water depth maps, also comprising of 50 members, is extracted from the pre-computed flood hazard datacube. An ensemble of binary maps is then generated where each pixel is classified either wet or dry, using a 10-cm threshold of water depth. The set of wet/dry pixels is then compared against the probabilistic flood map derived from a SAR image (in this study, from Sentinel-1 images processed by GFM algorithms). To optimally combine observations and simulations, we adopt a PF approach.
PFs have gained recent attention from the research community as they can hold promises for a fully non-linear DA and relax the assumption of Gaussian errors \citep{van2019particle}. In addition, the implementation of the SIS offers the advantage of preventing updates to model states that could potentially cause instabilities in the hydraulic model.
The prior and posterior probability distributions, representing model states across the model grid cells before and after an assimilation, are approximated using a set of \( N \) \textit{particles}. Each of them corresponds to the output from the forecast model (through GloFAS forecast and LISFLOOD-FP), with its own set of uncertainties and a \textit{weight}, which indicates the likelihood of that specific model output being \textit{correct} according to the observation.
During an assimilation, the PF analysis ascribes the particle weights based on flood information from the Sentinel-1-derived flood extent data resulted from GFM. 

The DA framework involves two key steps: the \textit{forecast} step, where model simulations are performed, and the \textit{analysis} step, during which the weights of $N$ particles $\lbrace x^1, x^2, ..., x^N\rbrace$ are updated based on available observations. The relationship between the observation vector $y$ (i.e. the observed flooded/non-flooded pixels) and the true state $x^t$ is expressed as $y = \mathcal{H}\left[x^t\right]+\epsilon$
where \(\mathcal{H}\) is the observation operator and $\epsilon$ represents observation errors.

At any given time step $k$, the prior PDF of the model state $x$ is described by a set of $N$ independent random particles $x^n$:
\begin{equation}
    p_k(x) \approx \sum_{n=1}^{N} \Theta_k^n \ \delta\left(x-x_k^n\right)
    \label{eq:dirac}
\end{equation}
where $\delta(z)$ is the Dirac delta function (which is zero everywhere except for $z$, its integral is equal to one), and the initial weights $\Theta_k^n$ are assumed to be uniform (i.e. $\Theta_k^n=1/N$, for $n=1,...,N$).

At the forecast step, the model (denoted by $\mathcal{M}$) propagates the particles without approximation:
\begin{equation}
    p_{k+1}(x) \approx \sum_{n=1}^{N} \Theta_k^n \ \delta\left(x-x_{k+1}^n\right)
    \textnormal{ with }
    x_{k+1}^{n} = \mathcal{M}_{k+1}\left(x_{k}\right)
\end{equation}

The analysis step updates the weight of each particle according to the likelihood of the particle given the observation:
\begin{equation}
    \Theta_{k+1}^{a,n} \propto \Theta_{k+1}^{f,n} \  p\left(y_{k+1} | x^n_{k+1}\right)
\end{equation}
where superscript $a$ stands for \textit{analysis} and superscript $f$ stands for \textit{forecast}.


Similar to previous works \citep{giustarini2011assimilating,hostache2018near}, here the likelihood (global weight, $\Theta^n$ for particle $n=1,..,N$) is computed by the product of the pixel-based likelihoods (local weights $\theta^n_{i}$ for pixel $i=1,..,L$ of member $n$), assuming the $L$ pixel observation errors to be independent from each other. At time $k$ of the observation, local weights $\theta^n_{i,k}$ are defined for each particle $n$ and for each pixel $i$ according to \citep{hostache2018near}:
\begin{equation}
    \theta^n_{i,k} = p_{i,k}( \mathcal{W} | \sigma_0) \times M^n_{i,k} + \lbrace 1-p_{i,k}( \mathcal{W} | \sigma_0)\rbrace \times (1-M^n_{i,k})
    \label{eq:theta_in}
\end{equation}
where \(p_i(\mathcal{W} | \sigma_0)\) represents the probability of a pixel being classified as \textit{wet} (\(\mathcal{W}\)) based on the Sentinel-1 observations, and \(\{1 - p_i(\mathcal{W} | \sigma_0)\}\) represents the probability of it being classified as \textit{dry}. \(M^n_{i,k}\) is set to "1" if the model predicts the pixel as \textit{wet}, and "0" if the model predicts it as \textit{dry}. The simulated water depth maps are transformed into binary flood extent maps by treating a pixel as wet if the water level exceeds 10 cm. Applying Equation \eqref{eq:theta_in} assigns higher probabilities to pixels where model predictions align with observations. Finally, \(\Theta^n\) is computed for each particle by the normalized product of local weights, ensuring that the sum of global weights equals 1, as follows, 
\begin{equation}
    \Theta^n = \dfrac{\prod_{i=1}^{L} \theta_{i,n}}{\sum_{n=1}^{N}\prod_{i=1}^{L} \theta_{i,n}}
    \label{eq:globalW}
\end{equation}

Additionally, unless the number of particles $N$ increases exponentially with the dimensionality of the system state, the PF is prone to degeneration \citep{kong1994sequential}, as high probability is often concentrated in a single particle, leaving the rest with minimal weights \citep{van2019particle}. PFs commonly face such degeneration issues, particularly when $N$ is insufficient due to computational constraints \citep{zhu2016implicit}. Following the application of a standard PF, weight variance typically increases, causing only a few particles to retain significant weight. To address this issue, \citep{dimauro2022tempered} proposed an adaptation of the global weight from Equation~\eqref{eq:globalW} using a tempering coefficient $\alpha$, as shown in the following Equation:
\begin{equation}
    \Theta_n(\alpha) = \dfrac{\prod_{i=1}^{L} \theta_{i,n}^\alpha}{\sum_{n=1}^{N}\prod_{i=1}^{L} \theta_{i,n}^\alpha}
    \label{eq:globalW_alpha}
\end{equation}

The resulting global weights are used to estimate the water levels ($H$) and discharge ($Q$) at time $k$ and per pixel $i$, as follows,
\begin{equation}
    \overline{H}_i = \sum_{n=1}^{N}\Theta^{n} \cdot H^n{i}
    \label{eq:hmean}
\end{equation}

\begin{equation}
    \overline{Q}_j = \sum_{n=1}^{N}\Theta^{n} \cdot Q^n_{j}
    \label{eq:Qmean}
\end{equation}


It should be noted that while the water levels $H$ are pixel-based following the datacube resolution, the discharge estimates $Q$ are only available at GloFAS grid cells.
The particles retain their global weights until the next assimilation step. Afterward, all particles are reset to an equal weight before a new analysis step begins. In contrast, the expectation in the Open Loop (OL), i.e. without DA, is simply the ensemble mean, as each particle holds the same relative importance.

\section{RESULTS}\label{sec:results}
\subsection{Assessment metrics}
\subsubsection{1D metrics for water level and discharge time-series assessment}\label{sect:metrics}

The quality of the simulated water level $H^m$ (respectively, discharge $Q^m$) is assessed with respect to in-situ observations $H^o$ (respectively, discharge $Q^o$) in terms of root-mean-square error ($\mathrm{RMSE}$), Kling-Gupta efficiency ($\mathrm{KGE}$) \citep{gupta2009decomposition} and Nash–Sutcliffe model efficiency coefficient ($\mathrm{NSE}$) \citep{nash1970river} over the flood event. $\mathrm{RMSE}$ is computed between the simulated and the observed WL time-series, sampled at observation times, for the entire flood event: 

\begin{equation}\label{eq:RMSE}
 \mathrm{RMSE} = \sqrt{\dfrac{1}{n_{obs}} \sum_{i=1}^{n_{obs}}(H_i^m-H_i^o)^2}   
\end{equation}

On the other hand, $\mathrm{NSE}$ reflects on the ratio of the error variance of the simulated time-series divided by the variance of the observed time-series:
\begin{equation}
\mathrm{NSE}=1-\dfrac{\sum_{i=1}^{n_{obs}}(H_i^m-H_i^o)^2}{\sum_{i=1}^{n_{obs}}(H_i^o-\overline{H^o})^2}
\end{equation}
where $\overline{H^o}$ denotes the time-averaged observed water level over the event. For a perfect model, the estimation error variance computed with respect to observation is equal to 0 (i.e. $\sum_{i=1}^{n_{obs}}(H_i^m-H_i^o)^2 = 0$), thus the resulting $\mathrm{NSE}$ is equal to 1. 
Resulting $\mathrm{NSE}$ values nearer to 1 suggest a model with greater predictive capacity.
A model that produces an estimation error variance equal to the variance of the observed time-series results in a $\mathrm{NSE}$ equal to 0.
Furthermore, when the estimation error variance computed with respect to observation is larger than the variance of the observations, the $\mathrm{NSE}$ becomes negative.
In other words, a negative $\mathrm{NSE}$ occurs when the observed mean is a better predictor than the model.


Lastly, $\mathrm{KGE}$ is a statistical metric widely adopted in hydrology and water resources engineering to evaluate model performance in simulating streamflow, groundwater recharge, and water quality parameters. It was introduced to address the shortcomings of other metrics, such as the $\mathrm{KGE}$ and the coefficient of determination ($\mathrm{R}^2$), which primarily focus on replicating the mean and variance of observed data. The $\mathrm{KGE}$ combines three components: the Pearson correlation coefficient ($r$), the variability ratio ($\gamma$), and the bias ratio ($\beta$), providing a comprehensive evaluation of model performance: 
\begin{equation}
\mathrm{KGE}=1-\sqrt{\left[1-r(H^m,H^o)\right]^2 + \left[1-\beta(H^m,H^o)\right]^2+ \left[1-\gamma(H^m,H^o)\right]^2}
\end{equation}
\begin{equation}
r(H^m,H^o) = \dfrac{\mathrm{cov(H^o,H^m)}}{\mathrm{std}(H^o)\mathrm{std}(H^m)}
\end{equation}
\begin{equation}
\beta(H^m,H^o) = \dfrac{\mathrm{mean}(H^m)}{\mathrm{mean}(H^o)}
\end{equation}
\begin{equation}
\gamma(H^m,H^o) = \dfrac{\mathrm{std}(H^m)/\mathrm{mean}(H^m)}{\mathrm{std}(H^o)/\mathrm{mean}(H^o)}
\end{equation}\\
The $\mathrm{KGE}$ value ranges from negative infinity to 1, with 1 representing a perfect match between model predictions and observed data. It not only evaluates the accuracy of predictions but also considers the model’s ability to capture the variability and timing of observed data. 

\subsubsection{2D metrics for flood extent assessment}    
In order to assess the comparison between the simulated and the observed flood extents, the Critical Success Index ($\mathrm{CSI}$) is used.
$\mathrm{CSI}$ evaluates the simulated flood extent maps based on the ground truth that is the observed flood maps.
These indices are based on the count of pixels in one of four possible outcomes, forming a contingency map: True Positives ($TP$, or \textit{hits}) represent the number of correctly predicted flooded pixels, while True Negatives ($TN$) denote the correctly identified non-flooded pixels. False Positives ($FP$, or \textit{false alarms}), also known as \textit{over-prediction}, refer to non-flooded pixels incorrectly predicted as flooded, and False Negatives ($FN$, or \textit{misses}), also called \textit{under-prediction}, represent flooded pixels that were not correctly identified.

\begin{equation}
\mathrm{CSI}=\dfrac{TP}{TP+FP+FN}.
\label{eq:CSI}
\end{equation}
This metric ranges from 0\% when there is no common area (i.e. no agreement) between the simulated and the observed flood extents, and reach their highest value of 100\% when the prediction provides a perfect fit to the observed flood extents.

\subsection{Evaluation of LISFLOOD-FP hydraulic model component}


\subsubsection{Model performance over the calibrating 2003 flood event}

\begin{figure}[h]
    \centering
    \includegraphics[width=0.9\linewidth]{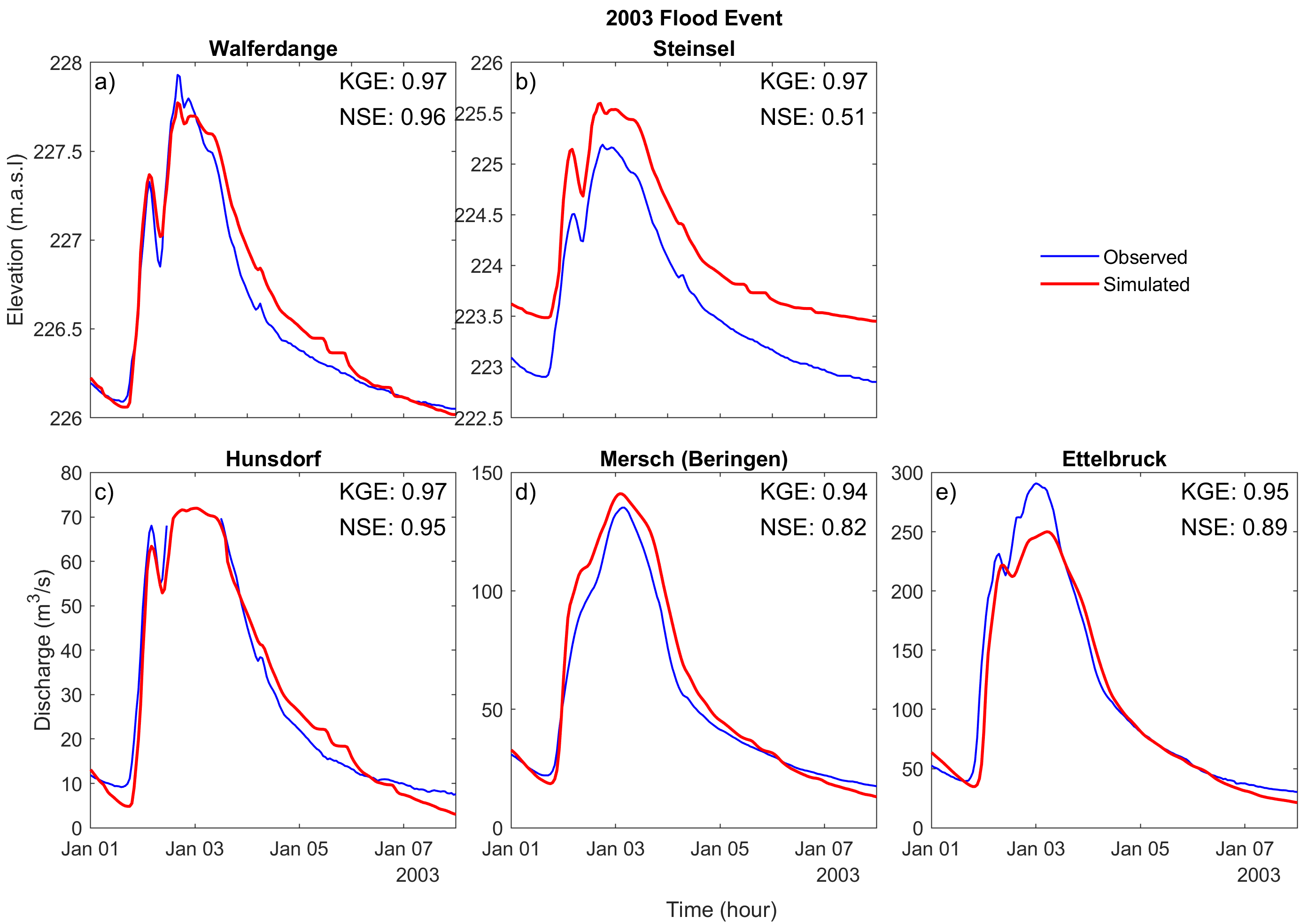}
    \caption{Observed (blue line) and simulated (red line) water levels during 2003 flood event at (a) Walferdange and (b) Steinsel stations, and discharges at (c) Hunsdorf, (d) Mersch, and (e) Ettelbruck stations.}
    \label{fig:calib_timeseries}
\end{figure}

It is worth-noting that an initial assessment of the model calibration revealed that the model tended to have over-prediction around the urban areas between Walferdange and Steinsel. In this regard, the model domain was divided into two regions, namely the first region between Pfaffenthal and Hunsdorf, and the second region between Hunsdorf and Ettelbruck, such that two different parameter sets were sampled in these two regions during the calibration process. 
Table~\ref{tab:NSE_KGE_2003} shows the 1D metrics $\mathrm{KGE}$ and $\mathrm{NSE}$ computed over the 2003 flood event between the simulated water levels/discharges and the observed ones at the gauge stations.
After the re-calibration, the model generally shows a satisfactory performance with $\mathrm{KGE}$ above 0.95 in all stations for the 2003 flood event. Figure~\ref{fig:calib_timeseries} reveals the simulated water levels at Walferdange and Steinsel, the simulated discharges at Hunsdord, Mersch (Beringen) and Ettelbruck in red lines, against the observed water levels/discharges in blue lines. It should be noted that the observed discharges at Hunsdorf station were discontinued for a couple of hours during the flood peak due to a technical problem. Nevertheless, the rising and falling limbs of all the hydrographs are well produced, capturing the timing of both flood onset and recession accurately. The peak discharges also show good agreement with the observed, despite a slight tendency of the model to overpredict at Steinsel and Mersch station. Overall, the LISFLOOD-FP model performs rather well across the entire domain, demonstrating its effectiveness in representing hydrological variations and dynamics.

\begin{table}[h]
    \centering
    \caption{$\mathrm{NSE}$ and $\mathrm{KGE}$ - 2003 flood event (used for calibration).}
    \label{tab:NSE_KGE_2003}
    \begin{tabular}{c|ccccccccc}
    \hline
    & Walferdange ($H$) & Steinsel ($H$) & Hunsdorf ($Q$) & Mersch ($Q$) & Ettelbruck / Alzette ($Q$)  \\\hline
    $\mathrm{KGE}$ & 0.97 & 0.97 & 0.97 & 0.94 & 0.95 \\ 
    $\mathrm{NSE}$ & 0.96 & 0.51 & 0.95 & 0.82 & 0.89\\\hline
\end{tabular}
\end{table}


\begin{figure}[h]
    \centering
    \includegraphics[width=0.5\linewidth]{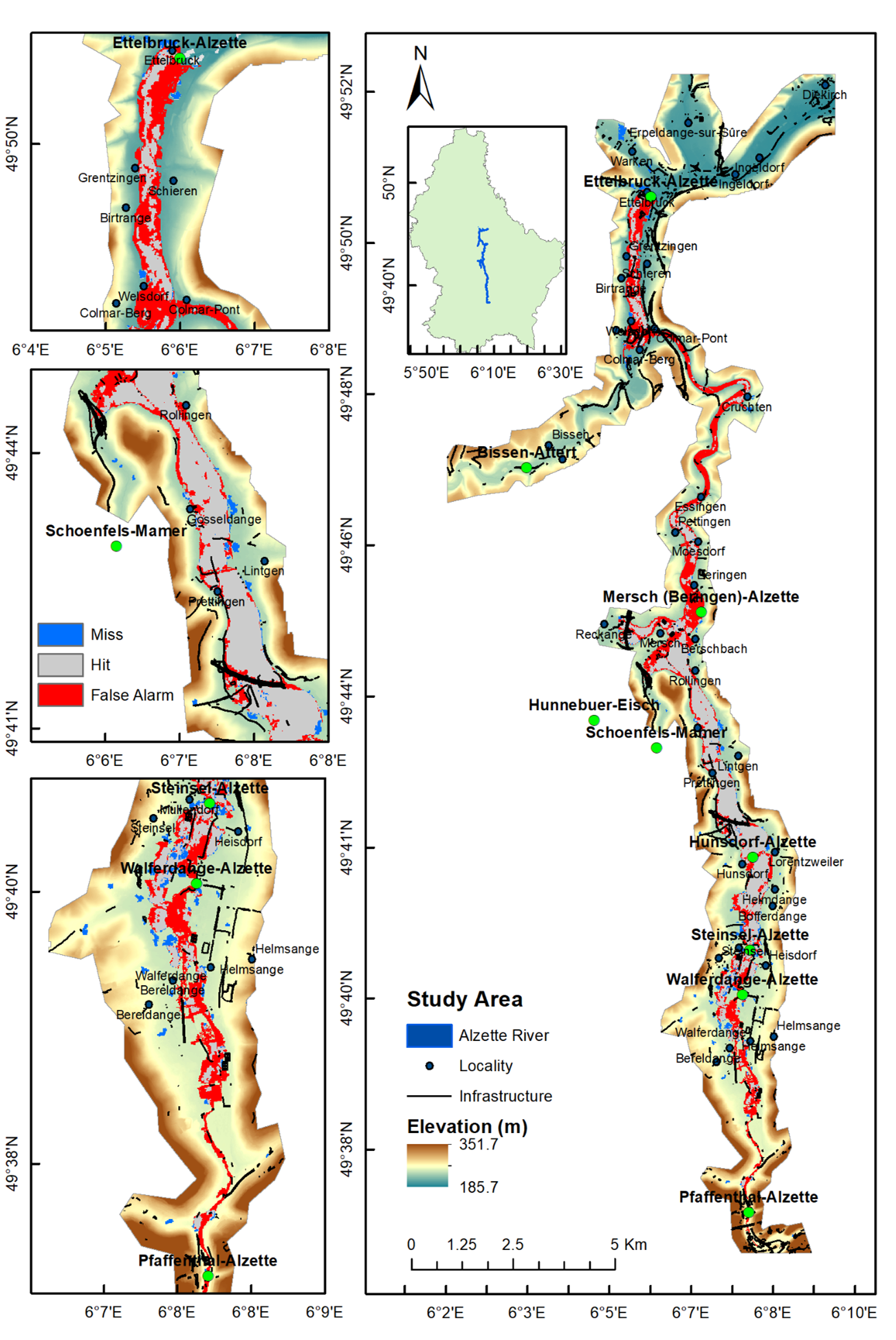}
    \caption{Model performance against the Sentinel-1 satellite imagery over the study reach in terms of contingency map, for the 2021 flood event.}
    \label{fig:calib_contingency}
\end{figure}

Figure~\ref{fig:calib_contingency} shows the contingency map between the simulated flood extent and Sentinel-1 observed flood extent maps, overlaid over the topography data.
\textit{Hit} (grey) means pixel is flooded in both simulation and observation, \textit{Miss} (blue) means pixel is flooded in observation but not in simulation, and \textit{False Alarm} (red) means pixel is flooded in simulation but not in observation. Pixel that is not flooded in both simulation and observation is not shown for visual clarity.


Evaluating against the satellite-derived water extent also shows some regions of over-prediction such as urban areas between Pfaffenthal and Steinsel, and river channel between Mersch and Ettelbruck (Figure~\ref{fig:calib_contingency}), result in a $\mathrm{CSI}$ of 0.49. However, it is acknowledged that these over-predictions could be due to the inability of Sentinel-1 to observe water body in narrow river channels and the difficulty in water detection in built-up areas. 

\subsubsection{Model validation against reference flood hazard maps}
Publicly available flood hazard maps based on the floods directive 2007/60/EC provide maps of flooded area of 10-year \footnote{\url{https://data.public.lu/fr/datasets/flood-hazard-maps-10-year-flood/}} and 100-year flood\footnote{\url{https://data.public.lu/fr/datasets/flood-hazard-maps-100-year-flood/}} events across Luxembourg. They were used here as supplementary information for quasi-validation. Since the flood hazard maps were generated using a different hydraulic modeling setup, they shall not be regarded as definitive references for rigorous validations. However, comparing the flood extent maps simulated by the calibrated model in this study to such existing flood hazard maps is still useful, as it helps ensure that the simulated flood extents are in reasonable agreement with the existing data. 

\begin{figure}[h]
    \centering
    \includegraphics[width=0.75\linewidth]{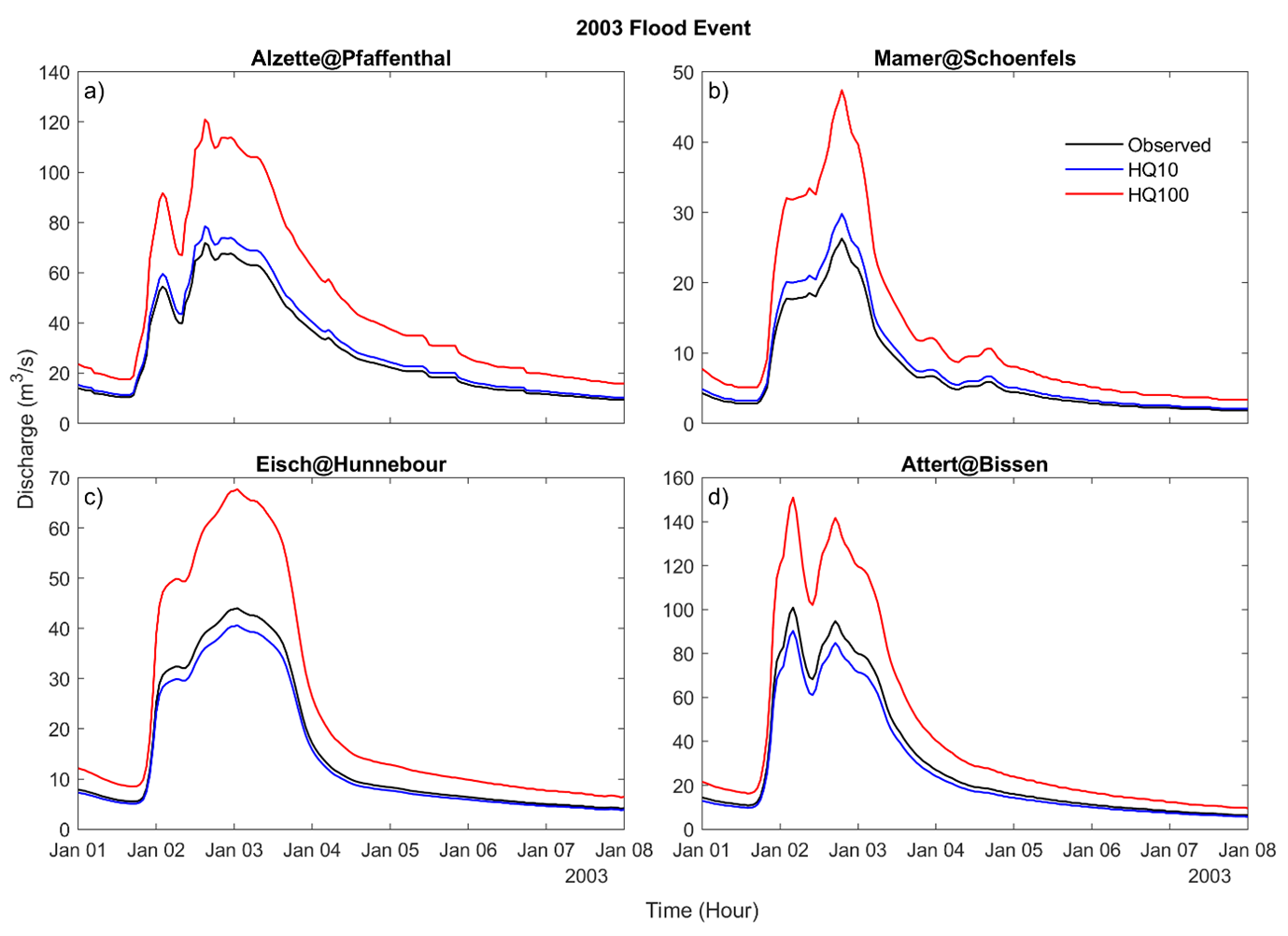}
    \caption{Hypothetical inflow discharge time-series based on the hydrograph of the 2003 flood event (black line) for the 10-year (HQ10) (blue line) and 100-year (HQ100) (red line) flood events at the four BCs: (a) Pfaffenthal, (b) Schoenfels, (c) Hunnebour, and (d) Bissen.}
    \label{fig:HQ10_HQ100}
\end{figure}

Accordingly, in order to carry out the quasi-validation, the calibrated model was configured as follows: 
\begin{enumerate}
    \item the peak discharge for the 10-year and 100-year flood events at the four gauging stations were obtained from the AGE (Administration de la Gestion de l'Eau), provided by Table~\ref{tab:return_period},
    \item due to a lack of information on the hydraulic modeling setup, it was assumed that the flood progression follows the shape of the January 2003 flood event, 
    \item the peak discharge obtained in 1) were used to scale the shape of the January 2003 flood event, generating hypothetical time series for the 10-year and 100-year flood events (blue and red lines in Figure~\ref{fig:HQ10_HQ100}, respectively), and
    \item these scaled hypothetical time series were used as upstream BCs for the calibrated model to produce the corresponding 10-year and 100-year flood extent maps. 
\end{enumerate}

\begin{figure}[h]
    \centering
    \begin{subfigure}{0.33\linewidth}
    \includegraphics[trim=0 0 18cm 0,clip,width=\linewidth]{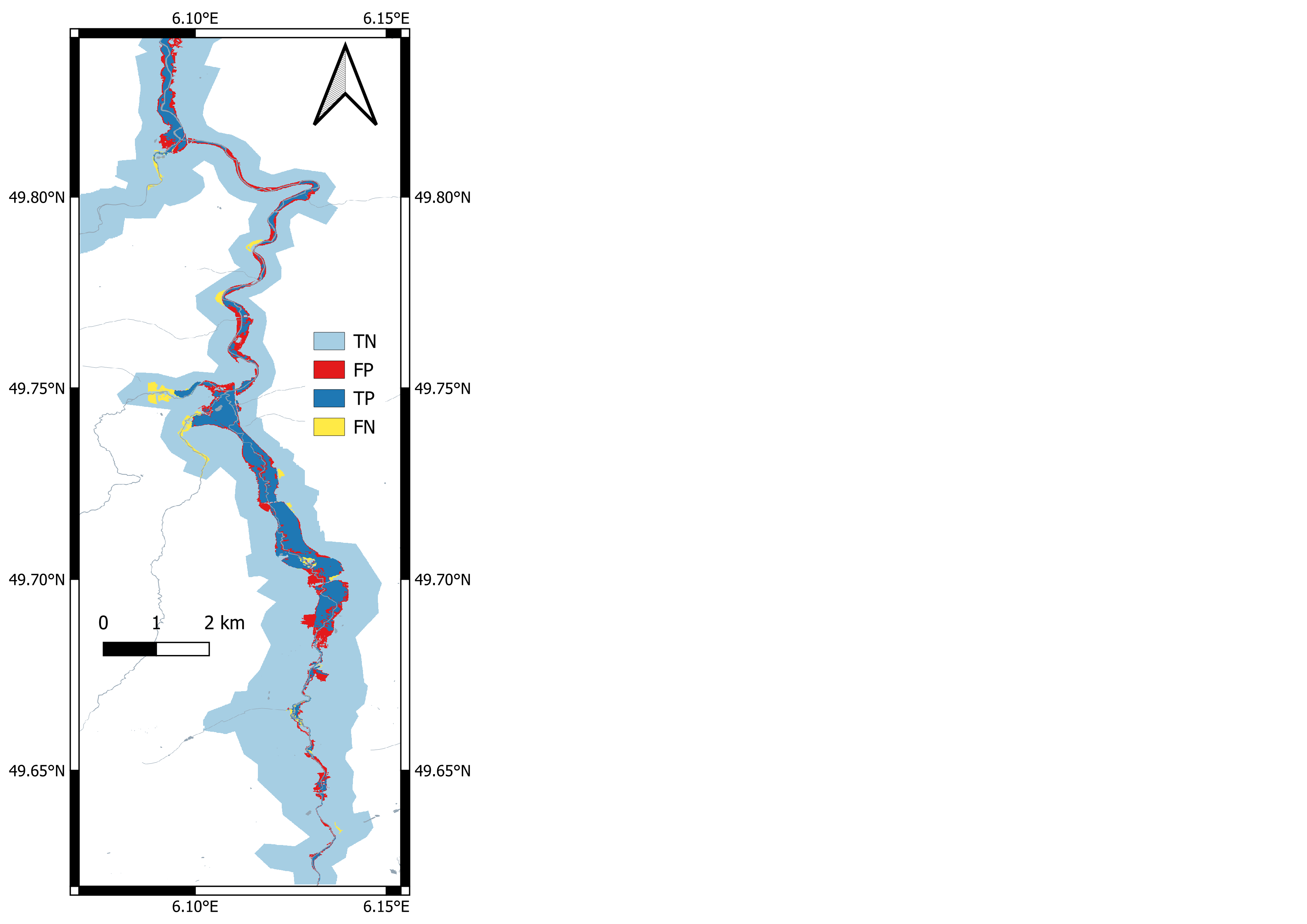}
    \caption{HQ10}
    \end{subfigure}
    \begin{subfigure}{0.33\linewidth}
    \centering
    \includegraphics[trim=0 0 18cm 0,clip,width=\linewidth]{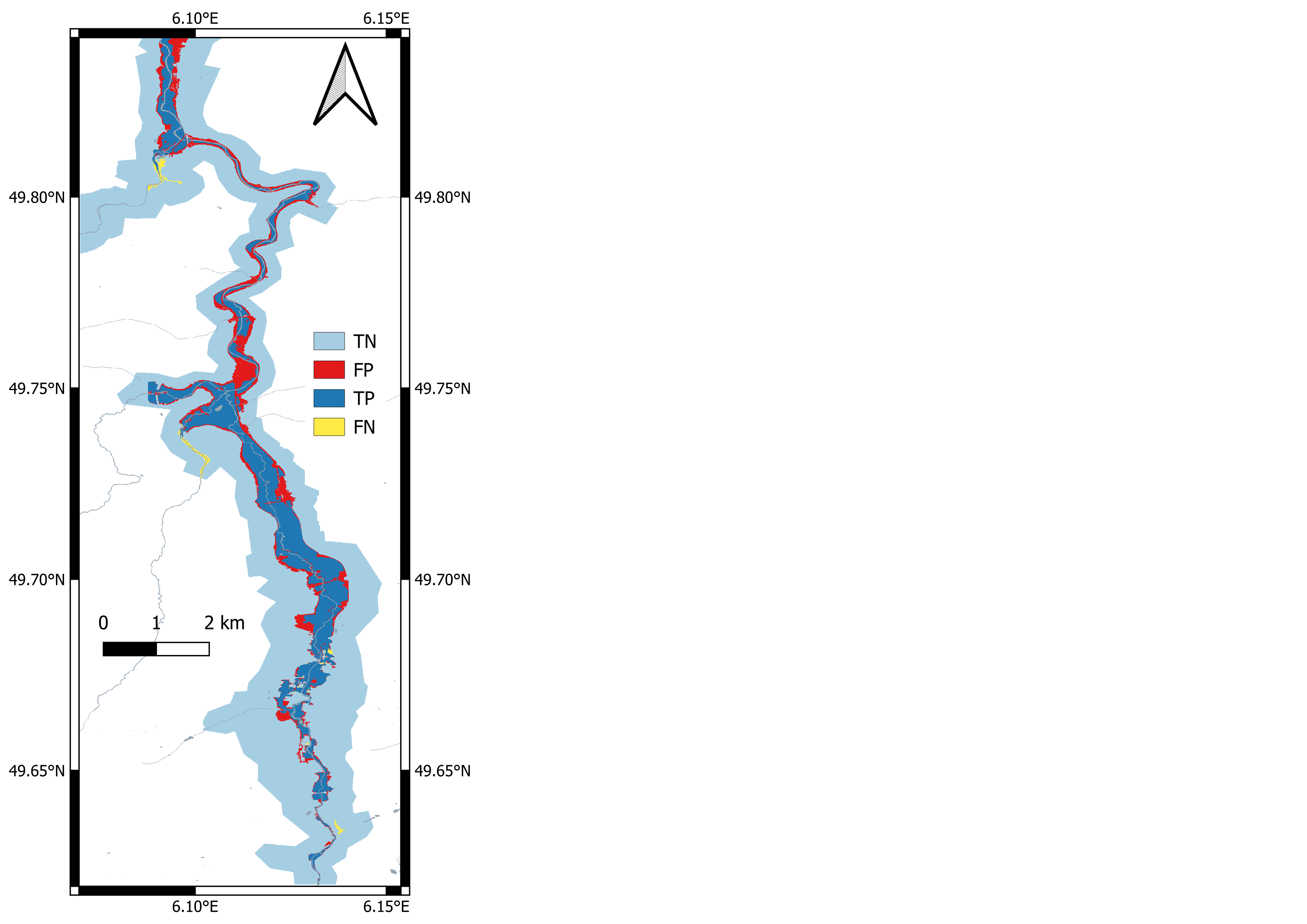}
    \caption{HQ100}
    \end{subfigure}
    \caption{Model performance against the flood extent map of \modif{geoportail} for: (a) the 10-year (HQ10) flood event, and (b) the 100-year (HQ100) flood event.}
    \label{fig:calib_contingency_HQ10_HQ100}
\end{figure}

Figure~\ref{fig:calib_contingency_HQ10_HQ100} depicts the contingency maps for the 10-year and 100-year simulated flood extent maps with respect to those of the \modif{geoportail}.
The pixels correctly predicted as flooded and non-flooded, respectively, are represented in dark blue and in light blue. 
The dry pixels mistakenly predicted as flooded (or \textit{over-prediction}) are indicated in red, whereas the wet pixels incorrectly predicted as non-flooded (or \textit{under-prediction}) are marked in yellow.
Validations against the flood extent maps of \modif{geoportail} (Figure~\ref{fig:calib_contingency_HQ10_HQ100}) show that the simulated flood extent has a very good agreement with that of \modif{geoportail}, reflecting the model is robust and has been well calibrated given all available observed data. Overall, the model performed quite well across the entire domain, demonstrating its effectiveness in representing hydrological variations and dynamics.


\subsubsection{Model performance over 2021 flood event}\label{ssec:assessment_2021}

Similar to Table~\ref{tab:NSE_KGE_2003}, Table~\ref{tab:NSE_KGE_2021} presents the 1D metrics $\mathrm{KGE}$ and $\mathrm{NSE}$ calculated for the 2021 flood event, comparing simulated water levels and discharges with the observed values at the gauge stations. Although these 1D metrics are slightly lower (i.e., reduced $\mathrm{KGE}$ and $\mathrm{NSE}$) compared to those in Table~\ref{tab:NSE_KGE_2003} from the calibration period, they remain within a high range, with $\mathrm{KGE}$ values all exceeding 0.82 and $\mathrm{NSE}$ values above 0.64. This is further supported by the visual comparison shown in Figure~\ref{fig:calib_timeseries}, which contrasts the simulated water levels at Walferdange and Steinsel, and the simulated discharges at Hunsdrof, Mersch (Beringen), and Ettelbruck (red lines), with the observed water levels and discharges (blue lines).

\begin{table}[h]
    \centering
    \caption{$\mathrm{NSE}$ and $\mathrm{KGE}$ - 2021 flood event.}
    \label{tab:NSE_KGE_2021}
    \begin{tabular}{c|ccccccccc}
    \hline
    & Walferdange ($H$) & Steinsel ($H$) & Hunsdorf ($Q$) & Mersch ($Q$) & Ettelbruck / Alzette ($Q$)  \\\hline
    $\mathrm{KGE}$ & 0.96 & 0.82 & 0.86 & 0.92 & 0.94 \\ 
    $\mathrm{NSE}$ & 0.88 & 0.54 & 0.62 & 0.70 & 0.84 \\\hline
\end{tabular}
\end{table}

\begin{figure}[h]
    \centering
    \includegraphics[width=0.9\linewidth]{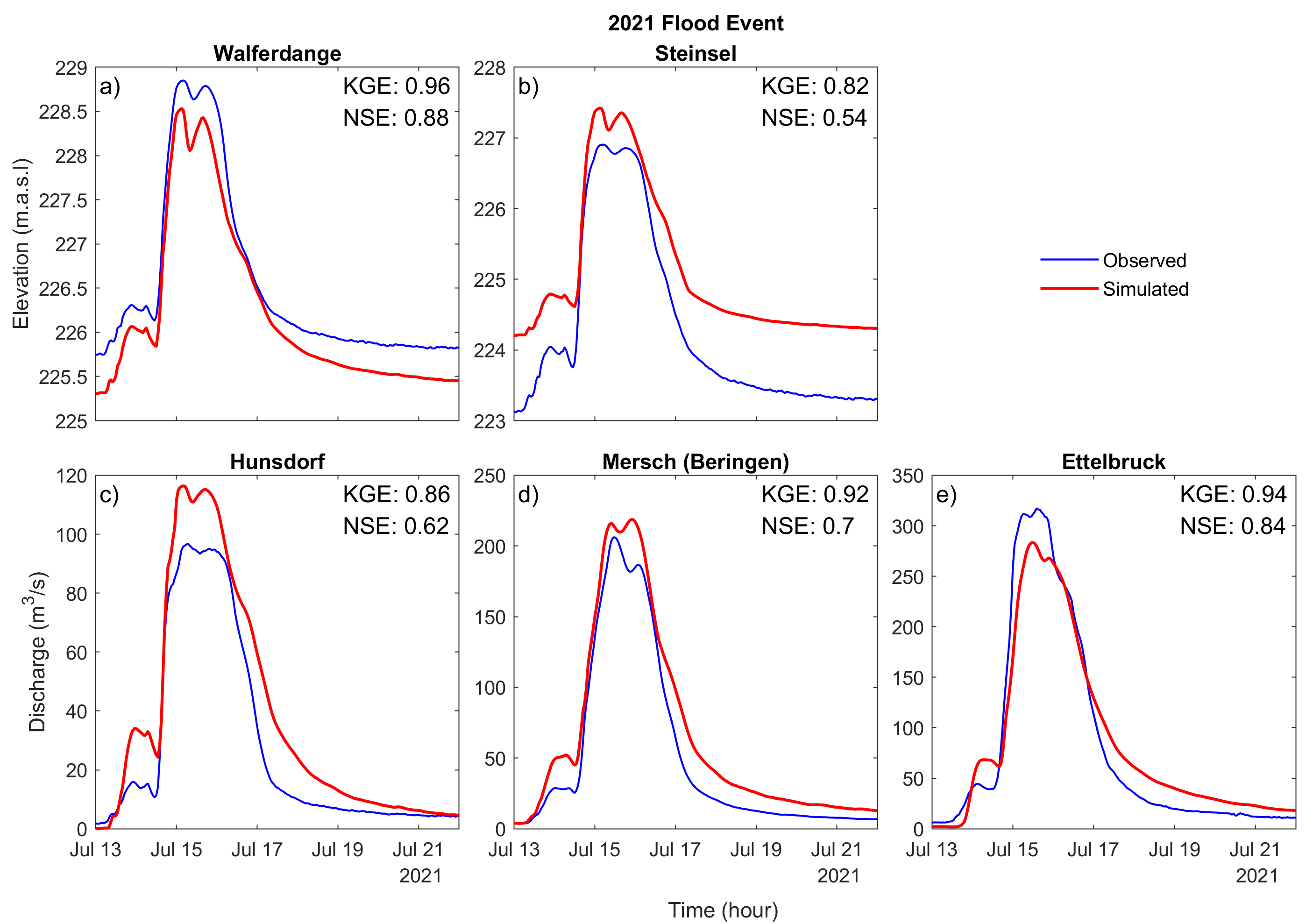}
    \caption{Observed (blue line) and simulated (red line) 
     water levels during 2021 flood event at (a) Walferdange and (b) Steinsel stations, and discharges at (c) Hunsdorf, (d) Mersch, and (e) Ettelbruck stations.}
    \label{fig:calib_timeseries_2021}
\end{figure}

\begin{figure}[h]
    \centering
    \begin{subfigure}{0.45\textwidth}
        \centering
        \includegraphics[trim=0 0.51cm 20.3cm   0,clip,width=0.45\linewidth]{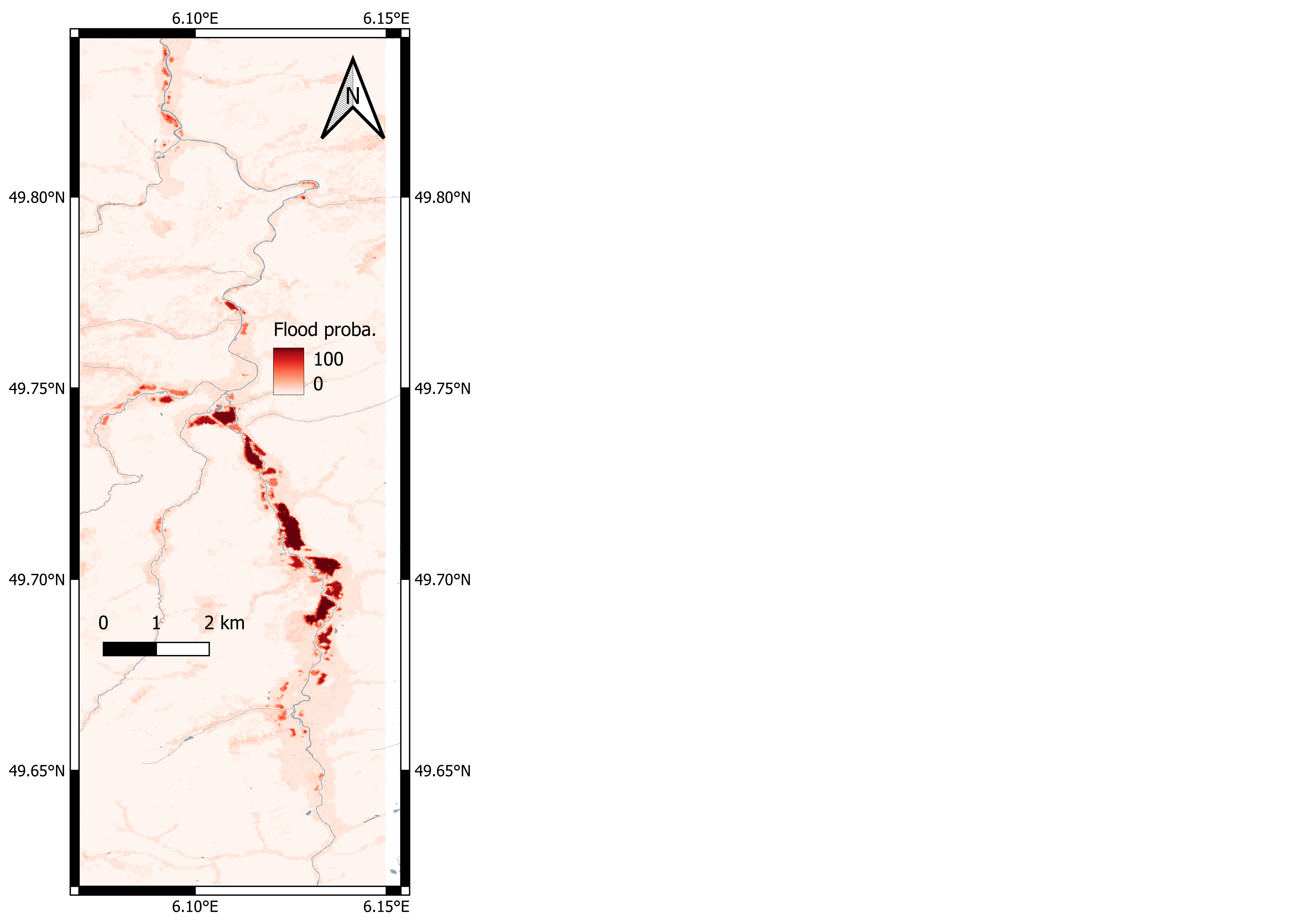}
        \includegraphics[trim=0 0.51cm 20.3cm   0,clip,width=0.45\linewidth]{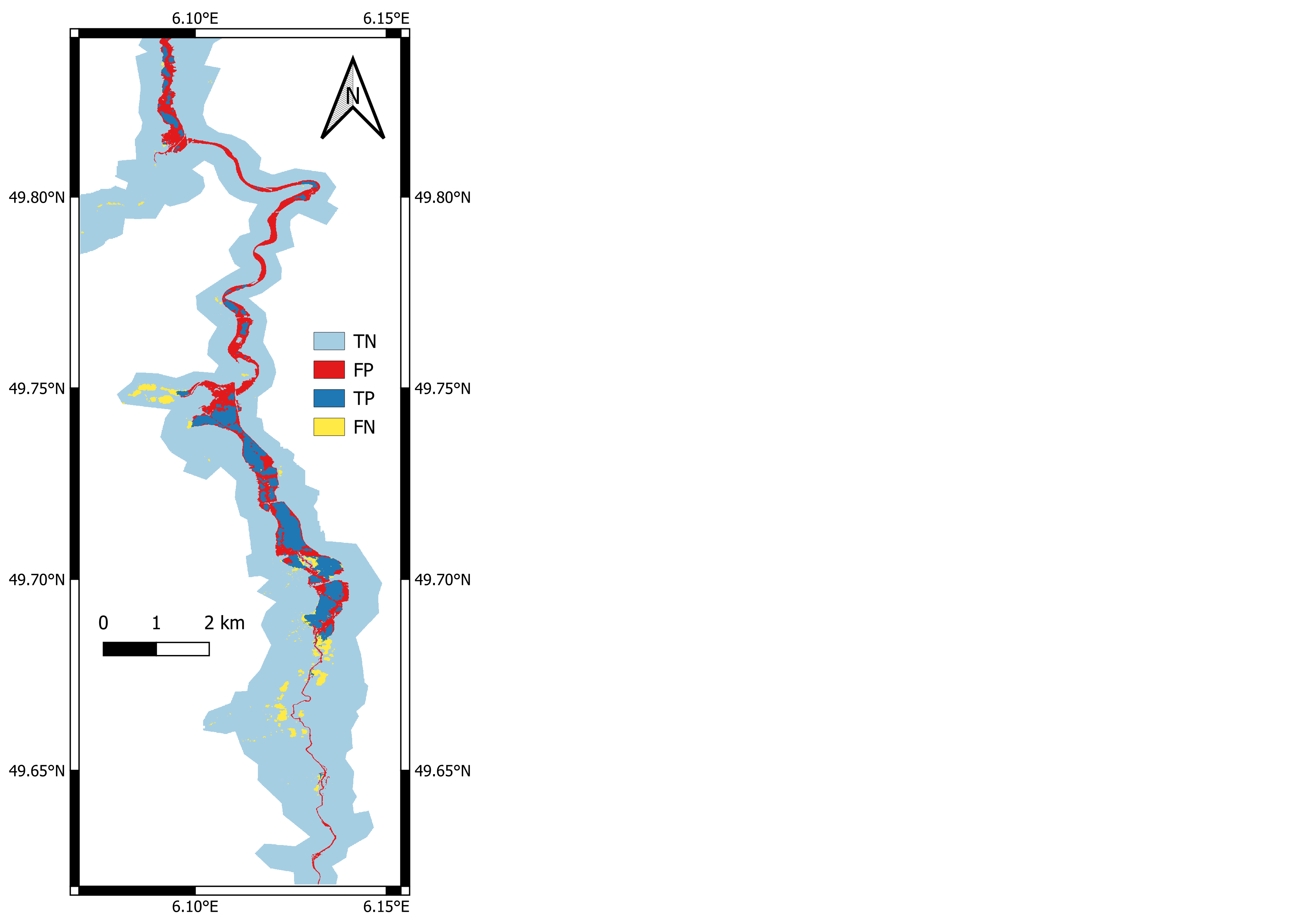}
        \caption{July 15}
    \end{subfigure}
    \hfill
    \begin{subfigure}{0.45\textwidth}
        \centering
        \includegraphics[trim=0 0.51cm 20.3cm   0,clip,width=0.45\linewidth]{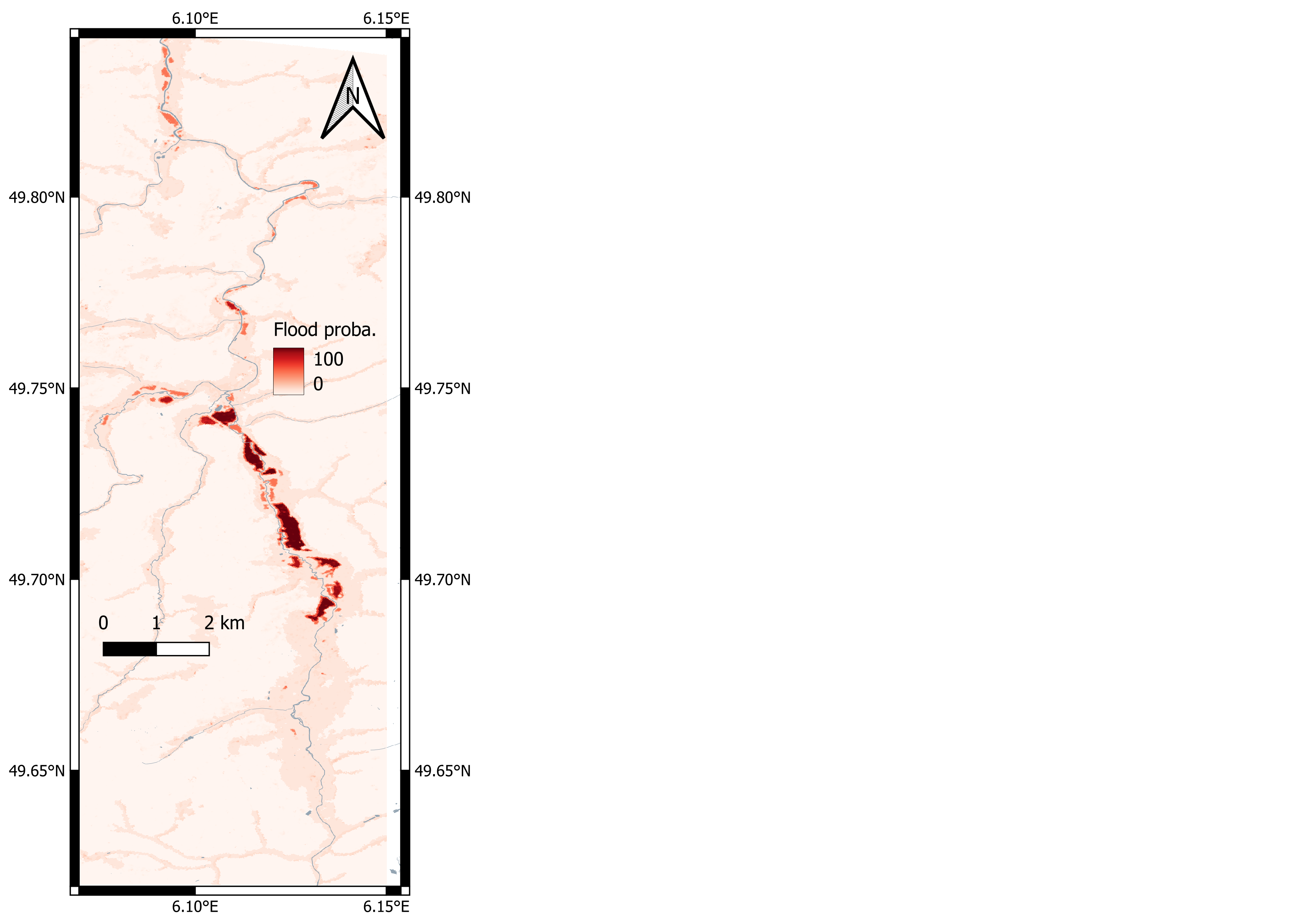}
        \includegraphics[trim=0 0.51cm 20.3cm   0,clip,width=0.45\linewidth]{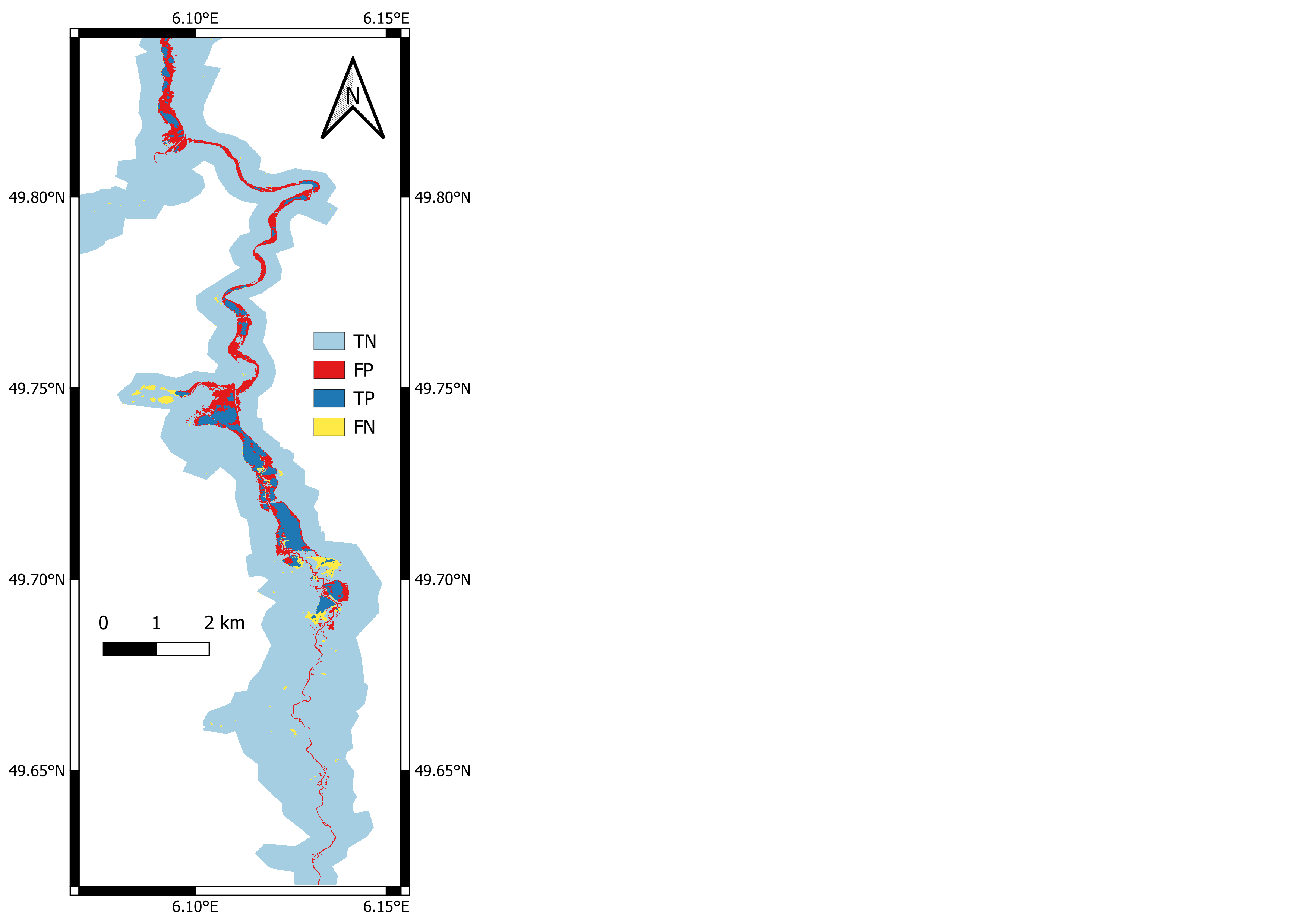}
        \caption{July 16}
    \end{subfigure}
    \caption{LISFLOOD-FP hydraulic model performance against the Sentinel-1-derived flood extent map observed on: (a) 2021-07-15 05:50:52Z, and (b) 2021-07-16 05:42:17Z. Left figure: Sentinel-1-derived flood probability maps from GFM; Right figure: contingency maps between simulated flood maps and observed flood maps.}
    \label{fig:contingency_2021}
\end{figure}

The flood peak of the 2021 event was captured by Sentinel-1 on its descending orbit 37 on July 15, 2021, at 05:50:52Z, and again the following day on descending orbit 139 at 05:42:17Z on July 16, 2021. Figure~\ref{fig:contingency_2021} shows the resulting flood probability maps (left panel) generated by GFM from these Sentinel-1 images, as well as contingency maps comparing the flood probability maps (using a 25\% flood probability threshold) with the simulated flood extent maps from the LISFLOOD-FP model. It's important to note that permanent water bodies are excluded from the GFM flood probability maps, which results in rivers and tributaries appearing as over-predicted areas.
Overall, the central region between Steinsel and Mersch (Beringen) shows strong agreement between the simulated and observed flood extents. On July 15, the hydraulic model displayed several areas of under-prediction (marked in yellow) between Pfaffenthal and Steinsel, which were no longer present on July 16. However, these areas are categorized as low certainty in the GFM flood probability maps.
Additionally, significant over-prediction (marked in red) is observed in the region between Mersch (Beringen) and Ettelbruck on both dates. This over-prediction is partly attributed to dense vegetation along the Alzette River, where the Sentinel-1 SAR signal is unable to detect floodwaters beneath the canopy. Such a situation (of reference/observed flood maps not showing floodwater) was not found in Figure~\ref{fig:calib_contingency_HQ10_HQ100} where the 10-year and 100-year flood hazard maps were based on hydraulic simulations.

\subsection{Evaluation of Sentinel-1-based DA framework}

At the core of this proposed Digital Twin dedicated to flood forecasting is the use of PF, a DA technique that enables dynamic model updates based on observational data. In this study, we assess the performance of assimilating Sentinel-1 images using two approaches: GFM and WASDI-based urban flood mapping, both within the PF framework. These correspond to two distinct experiments, referred to as PF1 and PF2, respectively. It is important to note that the OL experiment represents a forecast that is the mean of the 50 particles.

\begin{table}[h]
    \centering
    \caption{$\mathrm{RMSE}$ [m] of forecast water level over 30 days at gauge station Hunsdorf and Steinsel. The best (lowest) $\mathrm{RMSE}$ is boldfaced. \wdmax}
    \label{tab:rmse_H}
    \begin{tabular}{c|ccc|ccc|ccc}
    \hline
    & \multicolumn{9}{c}{GloFAS forecast issue date} \\
    \hline
    & \multicolumn{3}{c|}{July 12} & \multicolumn{3}{c|}{July 13} & \multicolumn{3}{c}{July 14} \\\hline 
    Exp. & OL & PF1 & PF2 & OL & PF1 & PF2& OL & PF1 & PF2 \\\hline 
    Hunsdorf & 0.179 & \textbf{0.151} & 0.151 & 0.191 & \textbf{0.127} & 0.127 & 0.206 & \textbf{0.143} & 0.144 \\\hline
    Steinsel & 0.181 & 0.141 & \textbf{0.139} & 0.145 & 0.106 & \textbf{0.104} & 0.130 & 0.103 & \textbf{0.102} \\\hline
    \end{tabular}
\end{table}


\begin{table}[h]
    \centering
    \caption{$\mathrm{NSE}$ [-] of forecast discharge over 30 days at gauge station Pfaffenthal (upstream BC) and Ettelbruck (downstream BC). The best (highest) $\mathrm{NSE}$ is boldfaced. \wdmax}
    \label{tab:rmse_Q}
    \begin{tabular}{c|ccc|ccc|ccc}
    \hline
    & \multicolumn{9}{c}{GloFAS forecast issue date} \\
    \hline
    & \multicolumn{3}{c|}{July 12} & \multicolumn{3}{c|}{July 13} & \multicolumn{3}{c}{July 14} \\\hline
    Exp. & OL & PF1 & PF2 & OL & PF1 & PF2& OL & PF1 & PF2 \\\hline
    Pfaffenthal & 0.618 & 0.711 & \textbf{0.718} & 0.551 & 0.752 & \textbf{0.760} & 0.587 & 0.763 & \textbf{0.767} \\\hline
    Ettelbruck & 0.570 & 0.663 & \textbf{0.669} & 0.509 & 0.661 & \textbf{0.671} & 0.631 & 0.740 & \textbf{0.743}\\\hline
    \end{tabular}
\end{table}


Table~\ref{tab:rmse_H} presents the $\mathrm{RMSE}$ values calculated over a 30-day period for the 2021 flood event, comparing the simulated water levels from the OL, PF1, and PF2 experiments with the observed water levels at the Hunsdorf and Steinsel gauge stations. These two stations were chosen because the flood extent assessment (subsection \ref{ssec:assessment_2021}) showed strong agreement between the model simulations and the observed flood maps from Sentinel-1. Similarly, Table~\ref{tab:rmse_Q} shows the $\mathrm{NSE}$s computed over time for the 2021 flood event, for the simulated inflow/outflow discharges from the OL, PF1, and PF2 experiments, with respect to the observed discharges at Pfaffenthal (the upstream BC) and Ettelbruck (the downstream BC). 
The PFs are applied using the GloFAS streamflow forecasts that issued on three  consecutive days, July 12, July 13 and July 14, to target the flood event that reached its peak on July 15.
For each observing station, the best metrics (i.e. lowest $\mathrm{RMSE}$ and highest $\mathrm{NSE}$) are boldfaced.
The quantitative assessments show that all three experiments, even the OL, yield accurate water level forecasts over the next 30 days. Indeed, the relatively low $\mathrm{RMSE}$ (under 20~cm) and high $\mathrm{NSE}$ (above 0.5) from the OL demonstrate the performance of the overall approach, even without the PF (that updates the weights among the particles), which combines the ensemble GloFAS streamflow forecasts and the hydraulic component. 

It is also shown that all DA experiments succeed in reducing the errors on the simulated water levels (i.e. $\mathrm{RMSE}$ under 15~cm) and discharge (i.e. $\mathrm{NSE}$ above 0.66) compared to those of OL. Indeed, the integration of satellite EO data (with Sentinel-1-derived flood extent maps) in both PF1 and PF2 allows for an average improvement of 15-33\% for $\mathrm{RMSE}$ at Hunsdorf and Steinsel, as well as an improvement of  15-36\% for $\mathrm{NSE}$ at Pfaffenthal and Ettelbruck by both PF1 and PF2. It should be noted that PF1 and PF2 yield very similar results. This is due the fact that only the flood extent maps on 2021-07-15 05:50:52Z presents a difference between the PF1 and PF2 with an overflow in urban areas, from which PF2 is more accurate than PF1 with GFM. 
The least improvements (from OL to PFs) were observed with only a 15\% reduction in $\mathrm{RMSE}$ and a 15\% increase in $\mathrm{NSE}$ when using the forecast issued on July 12 in both quantitative assessments. In contrast, the most significant improvements were achieved with the forecast issued on July 13. This suggests that using a two-day ahead forecast provides the most accurate results.

\begin{figure}[h]
    \centering
    \begin{subfigure}{0.48\linewidth}
        \includegraphics[width=0.95\linewidth]{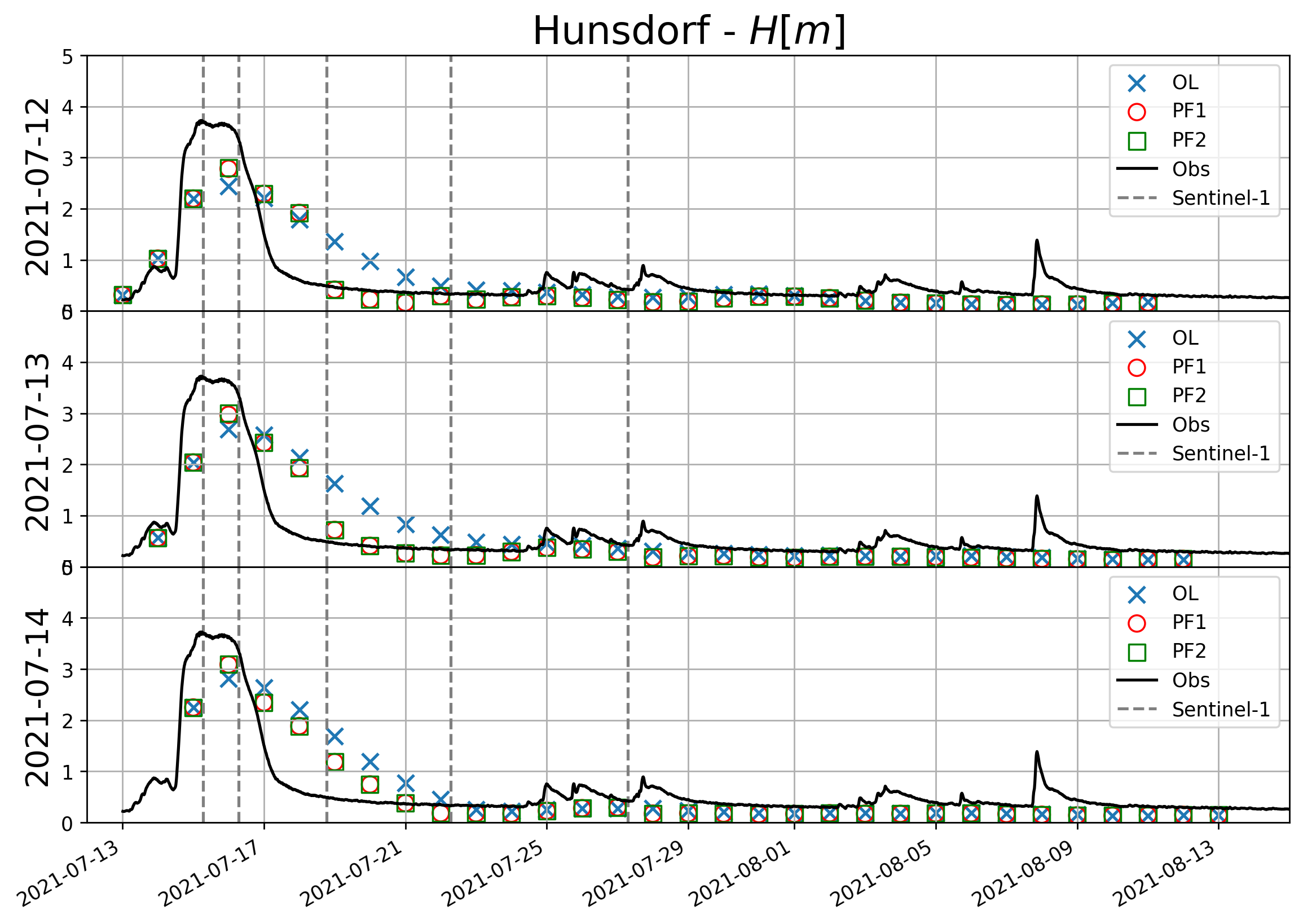}
        \caption{Hunsdorf}
        \label{}
    \end{subfigure}
    \hfill
    \begin{subfigure}{0.48\linewidth}
        \includegraphics[width=0.95\linewidth]{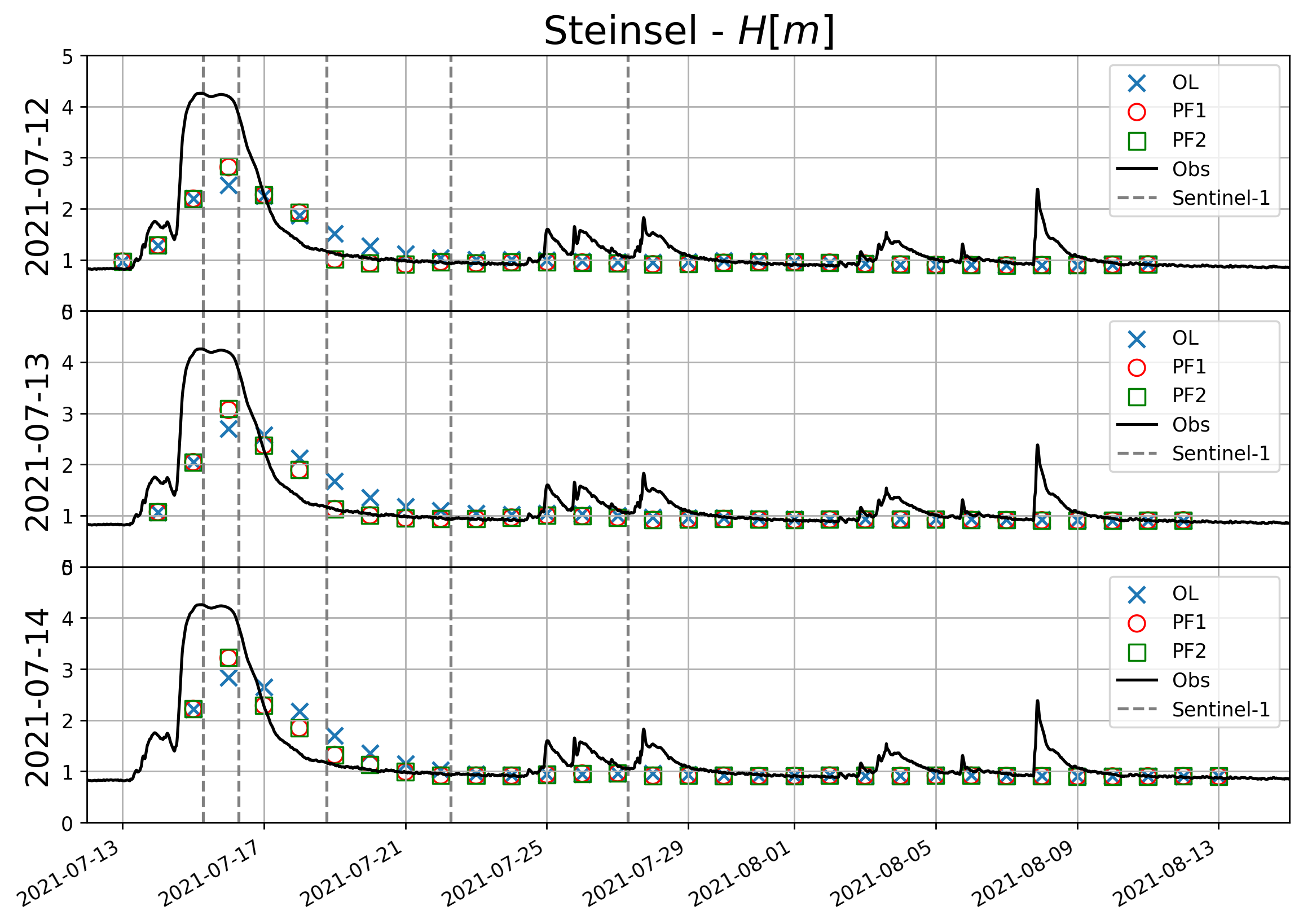}
        \caption{Steinsel}
        \label{}
    \end{subfigure}
    \caption{Forecast water levels over 30 days by OL (blue crosses), and by assimilation of GFM flood maps (red circles) and WASDI-based urban flood maps (green squares) compared to the observed water levels (black line) at (a) Hunsdorf and (b) Steinsel, using GloFAS streamflow forecasts issued on 2021-07-12 (top panels), 2021-07-13 (middle panels), and 2021-07-14 (bottom panels). Sentinel-1 overpass times are indicated as vertical dashed lines. \wdmax} 
    \label{fig:WL_assessment}
\end{figure}

\begin{figure}[h]
    \centering
    \begin{subfigure}{0.48\linewidth}
        \includegraphics[width=0.95\linewidth]{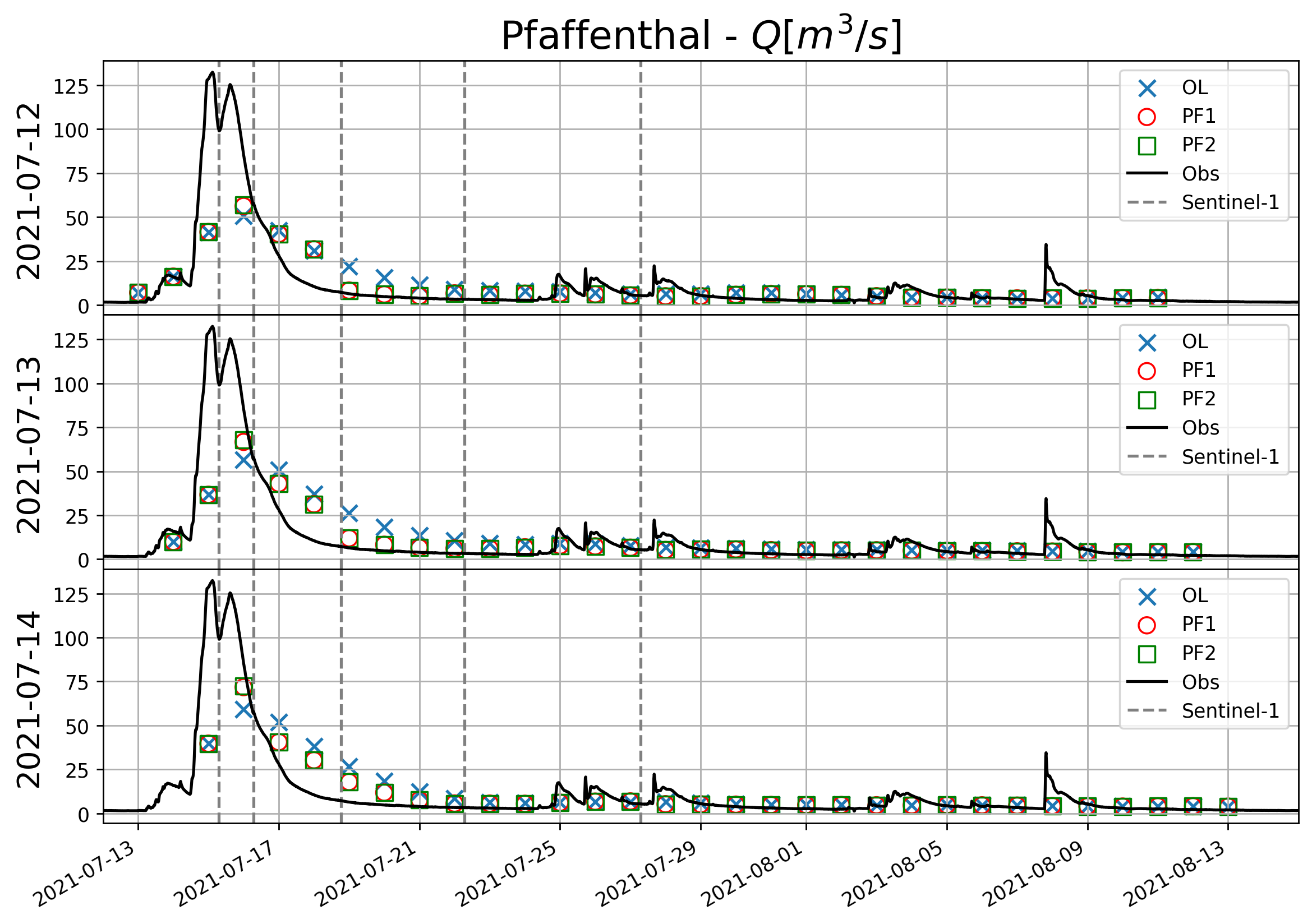}
        \caption{Pfaffenthal}
        \label{}
    \end{subfigure}
    \hfill
    \begin{subfigure}{0.48\linewidth}
        \includegraphics[width=0.95\linewidth]{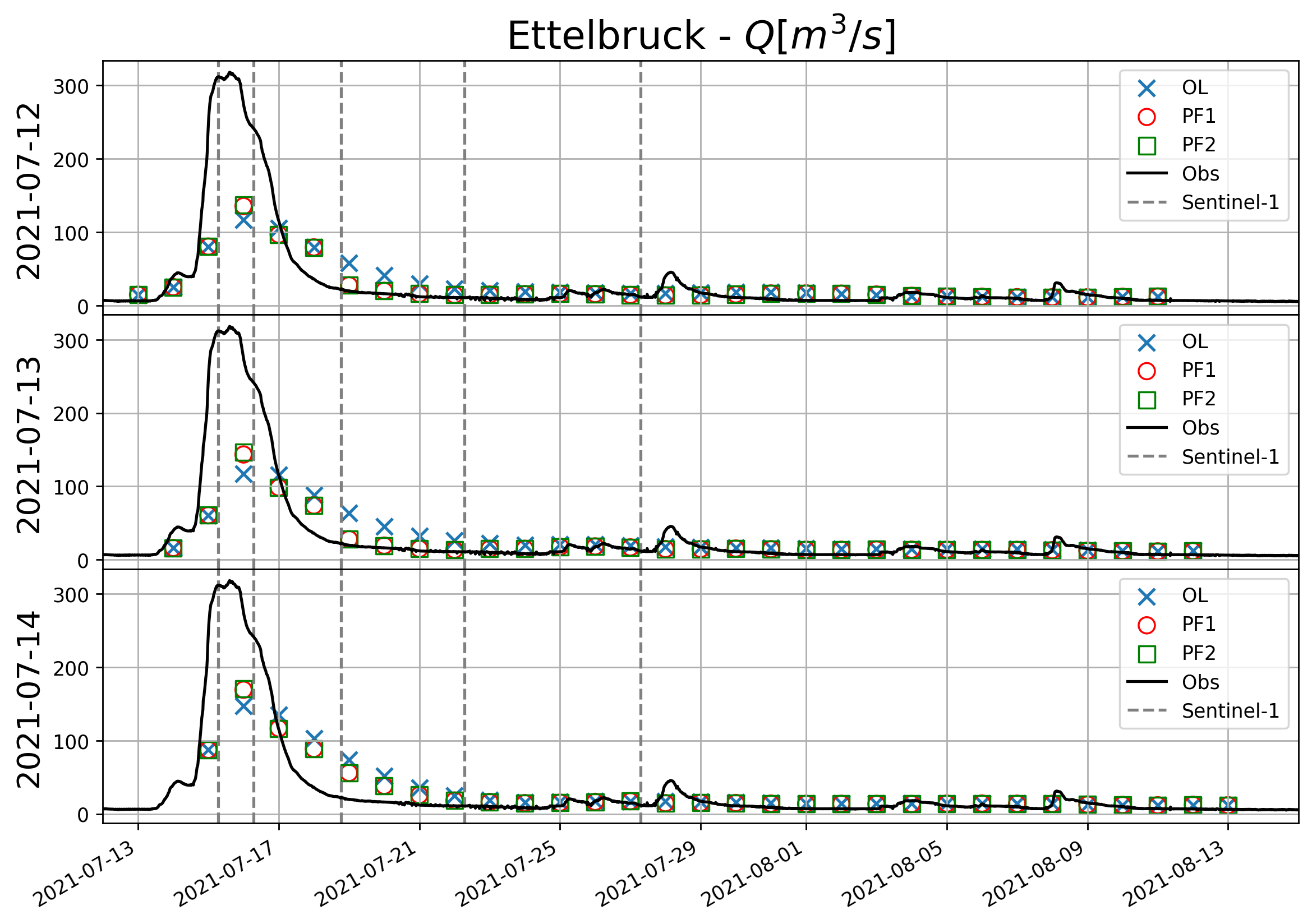}
        \caption{Ettelbruck}
        \label{}
    \end{subfigure}
    \caption{Forecast discharges over 30 days by OL (blue crosses), and by assimilation of GFM flood maps (red circles) and WASDI-based urban flood maps (green squares) compared to the observed discharges (black line) at (a) Pfaffenthal and (b) Ettelbruck, using GloFAS streamflow forecasts issued on 2021-07-12 (top panels), 2021-07-13 (middle panels), and 2021-07-14 (bottom panels). Sentinel-1 overpass times are indicated as vertical dashed lines. \wdmax} 
    \label{fig:Q_assessment}
\end{figure}

Figure~\ref{fig:WL_assessment} depicts the forecast water levels compared to the observed water level at observing stations: (a) Hunsdorf and (b) Steinsel, up to 30-day lead time; whereas Figure~\ref{fig:Q_assessment} reveals the forecast discharge at (a) Pfaffenthal and (b) Ettelbruck. The OL forecasts are depicted with blue crosses, while the PF1 and PF2 resulting forecasts are shown with red circles and green squares, respectively. The black lines represent the observed water levels and discharges the stations.
At each Sentinel-1 overpass, the PFs successfully update the particles to align with the observations. This is evident as the water levels and discharges, shown by red circles (PF1) and green squares (PF2), move closer to the observed water levels and discharges at the various monitoring stations (black lines). After each assimilation times, the PFs hold the same weights for the 50 particles until the next Sentinel-1 overpass. This results in the over-predictions of water levels and discharges after the flood peak until July 18 (third vertical dashed lines). 
However, it is worth noting that the flood peaks at all stations were under-predicted in all three experiments. This highlights a limitation of the proposed approach, primarily due to the inaccuracy of the forecast inflow discharges provided by GloFAS, as previously shown in Figure~\ref{fig:Q_GloFAS}. Additionally, the daily discharge data appears insufficient to produce accurate results in this case.

\begin{figure}[h]
    \centering

    \begin{subfigure}{0.19\linewidth}\includegraphics[trim=0 0.51cm 20.3cm   0,clip,width=\linewidth]{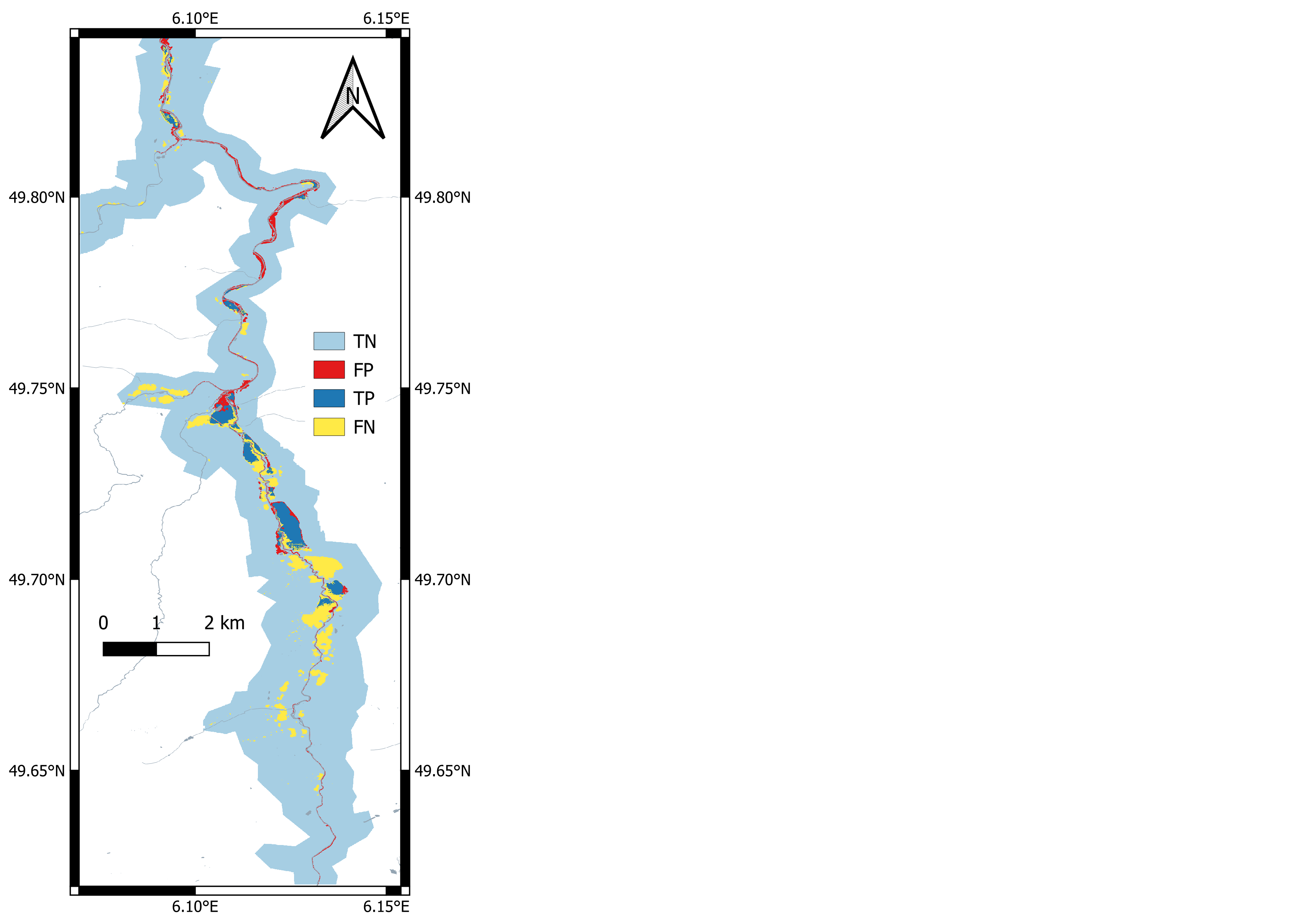}\caption{OL}\end{subfigure}
    \begin{subfigure}{0.19\linewidth}\includegraphics[trim=0 0.51cm 20.3cm   0,clip,width=\linewidth]{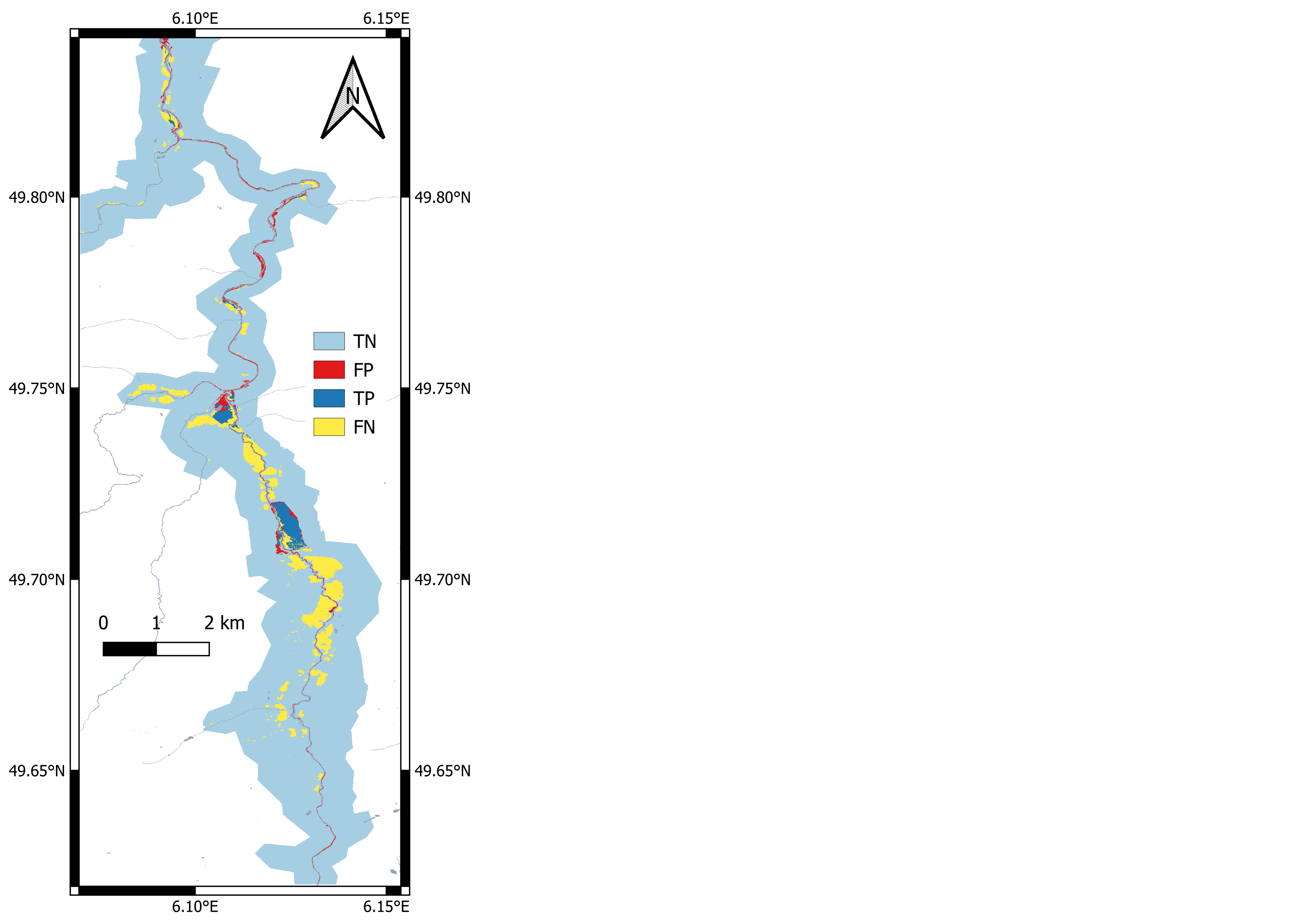}\caption{PF1}\end{subfigure}
    \begin{subfigure}{0.19\linewidth}\includegraphics[trim=0 0.51cm 20.3cm   0,clip,width=\linewidth]{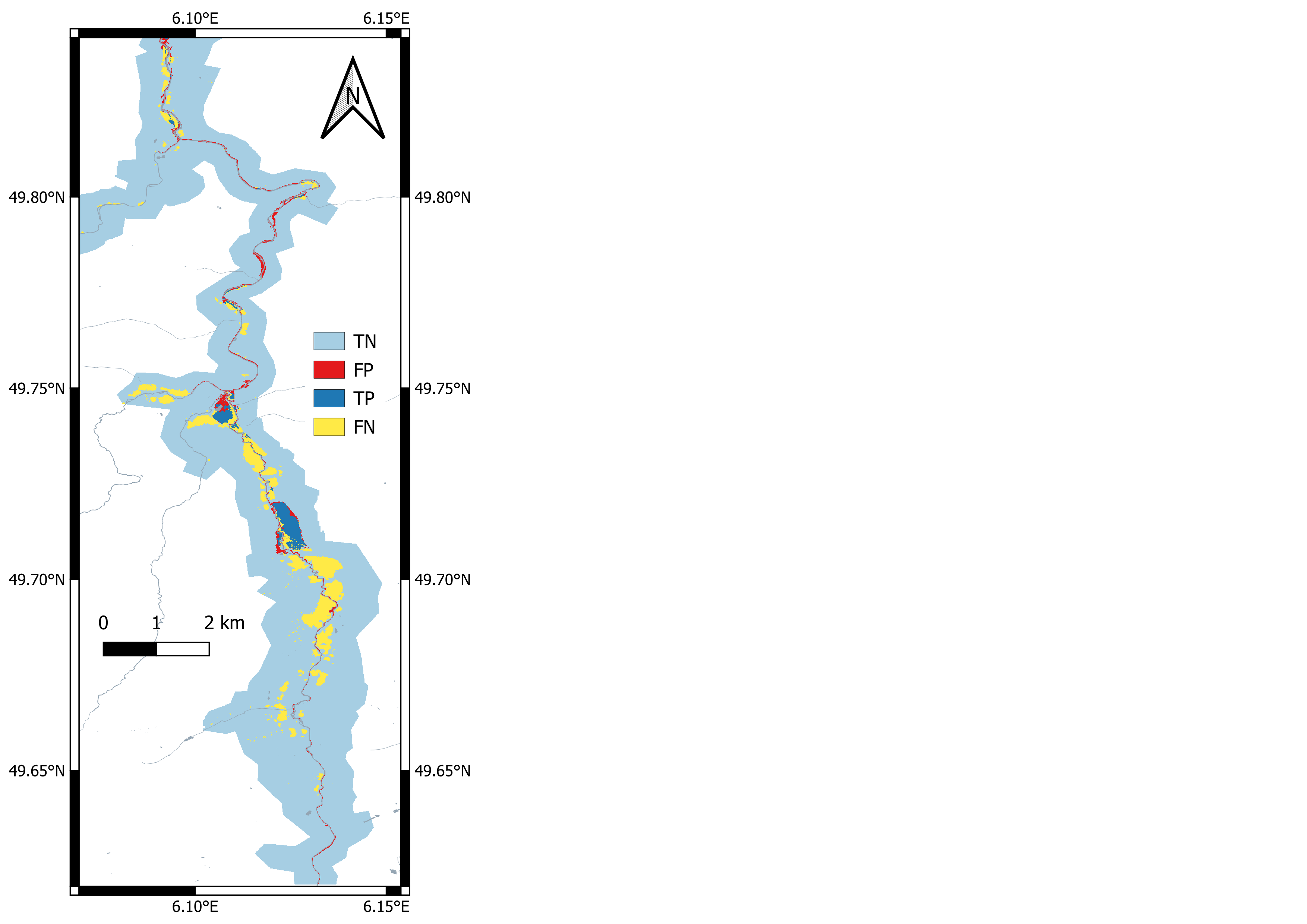}\caption{PF2}\end{subfigure}

    
    \caption{Assessment against Sentinel-1 flood extent map observed on 2021-07-15 05:50:52Z using GloFAS forecasts issued on 2021-07-13.} 
    \label{fig:contingency_2021_OL_PF}
\end{figure}

\begin{figure}[h]
    \centering
    
    \begin{subfigure}{0.19\linewidth}\includegraphics[trim=0 0.51cm 20.3cm   0,clip,width=\linewidth]{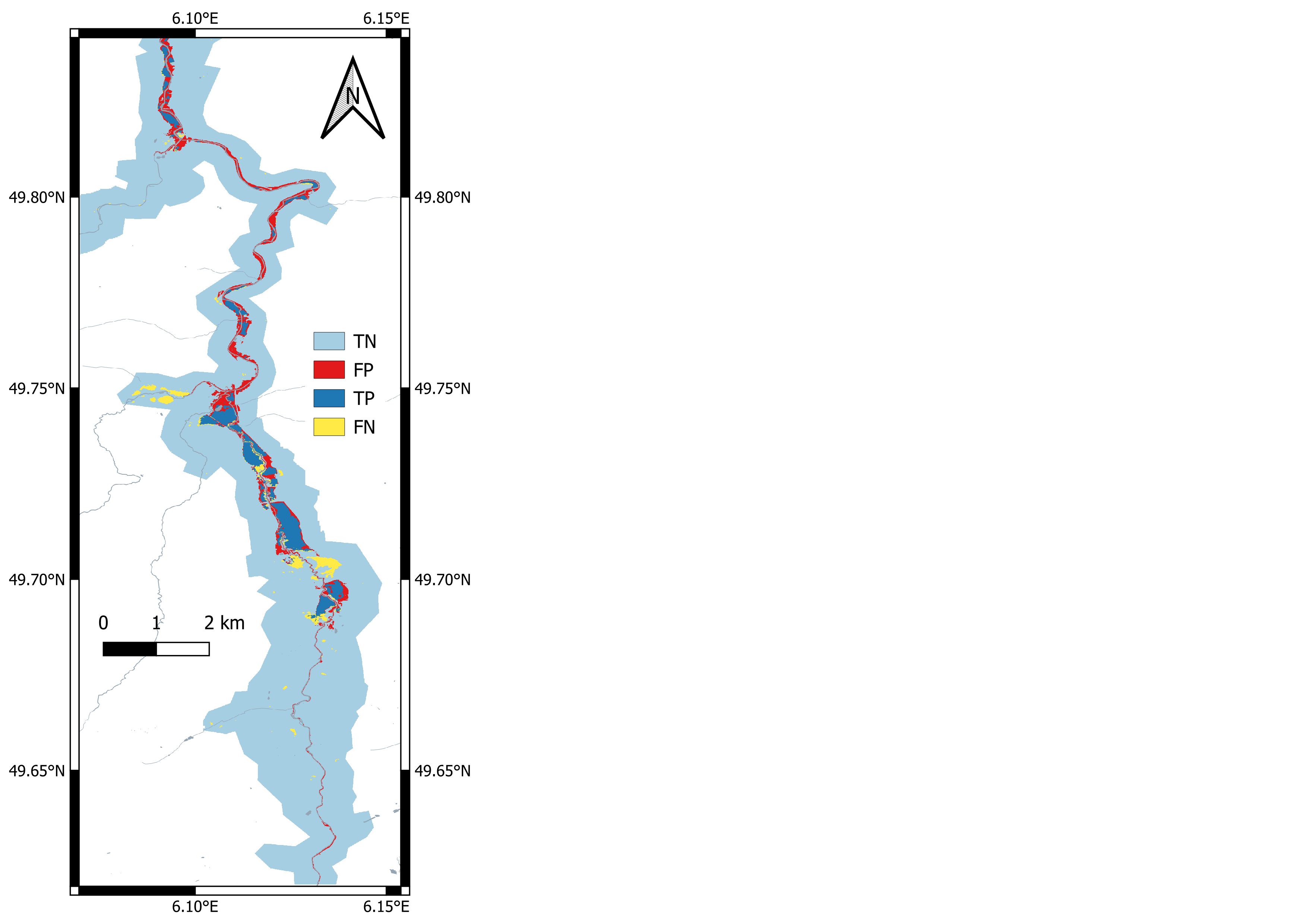}\caption{OL}\end{subfigure}
    \begin{subfigure}{0.19\linewidth}\includegraphics[trim=0 0.51cm 20.3cm   0,clip,width=\linewidth]{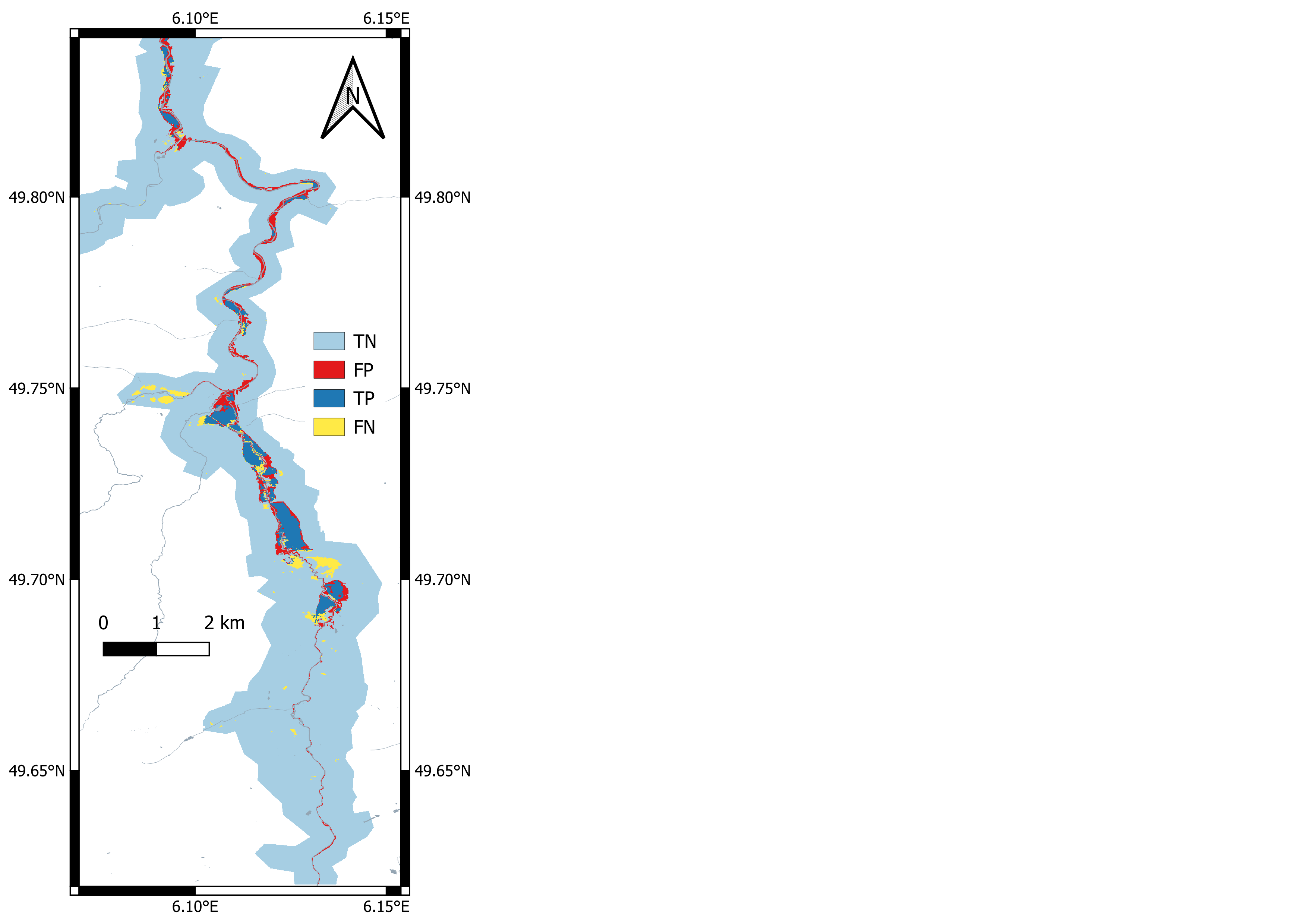}\caption{PF1}\end{subfigure}
    \begin{subfigure}{0.19\linewidth}\includegraphics[trim=0 0.51cm 20.3cm   0,clip,width=\linewidth]{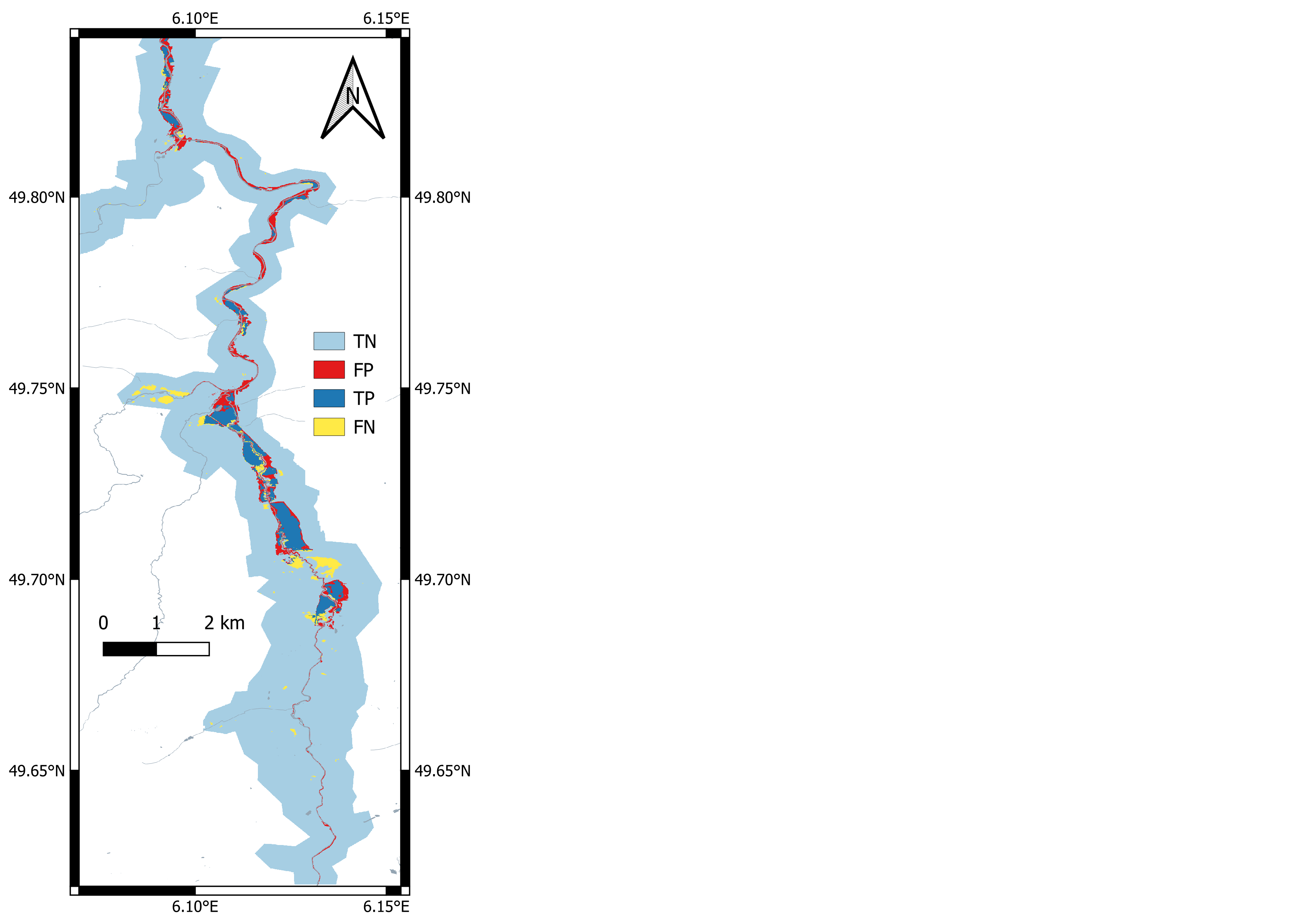}\caption{PF2}\end{subfigure}

    \caption{Assessment against Sentinel-1 flood extent map observed on 2021-07-16 05:42:17Z using GloFAS forecasts issued on 2021-07-13} 
    \label{fig:contingency_2021b_OL_PF}
\end{figure}

Similar to Figure~\ref{fig:contingency_2021}, Figure~\ref{fig:contingency_2021_OL_PF} and Figure~\ref{fig:contingency_2021b_OL_PF} show the contingency maps for the three forecast experiments compared to the Sentinel-1-derived flood maps on July 15, 2021, at 05:50:52Z, and on July 16, 2021, at 05:42:17Z, respectively. Here only the results using GloFAS forecasts issued on July 13 were shown.
It should be noted that, compared to model simulations using observed inflow discharges, the forecasts show significantly less agreement with the observed flood maps.
In contrast to the results from the previous 1D quantitative assessments, the contingency maps in Figure~\ref{fig:contingency_2021_OL_PF} from PF1 and PF2 are slightly worse than those from the OL, as shown by more under-predictions areas, while Figure~\ref{fig:contingency_2021b_OL_PF} shows very minimal differences between the contingency maps from OL and the two PFs. 
Such a difference is because while the OL uses an unweighted average, the PFs rely on a weighted average of the particles. The weight computation, however, was applied across the entire domain, which introduced errors, particularly due to misdetection in vegetated areas. As a result, the PFs might have favored incorrect scenarios. This highlights the need for a more localized weight computation, focusing only on the most reliable flooded areas. The use of exclusion maps \citep{zhao2021deriving} will be undertaken in future works to address this issue. 


\clearpage
\section{Discussions}

\subsection{Uncertainty of GloFAS streamflow forecast}
The uncertainty of GloFAS streamflow forecasts plays a significant role in influencing the accuracy of flood predictions. To provide medium- and extended-range discharge forecasts used in this research work, GloFAS relies on meteorological inputs, which inherently carry uncertainties. These uncertainties tend to be more pronounced at longer lead times, where the spread of forecasted discharge across ensemble members increases, as shown in Figure~\ref{fig:Q_GloFAS}. \citet{9553312} asserted that the underestimation by GloFAS is partly due to its design, which focuses on capturing flood events in larger rivers with upstream areas greater than 5,000 km$^2$.

The daily discharge of GloFAS averaged over the last 24 hours is also a limitation for real-time flood forecasting since it smooths out variations in river discharge, potentially missing crucial short-term fluctuations. Flood events often involve rapid changes in river discharge that can occur within hours, not days. By using discharge estimates averaging over a 24-hour period, critical peaks in discharge that may signal imminent flooding could be missed or delayed in the data. 

It is important to note that starting the forecast too early is not advisable due to the inaccuracy of the meteorological inputs in GloFAS. Therefore, for a flood event, initiating the forecast 2-3 days before the event would be the most effective approach. In addition, t is also advisable to adapt the proposed approach to the newer version 4.0 of GloFAS streamflow forecast with a spatial resolution of $0.05^\circ \times 0.05^\circ$.

\subsection{Generation of flood hazard catalog based on a specific past flood event}

In this research work, the flood hazard catalog has been constructed using the 2021 flood event as the input time-series to scale up the different scenarios. While this approach provides a concrete and detailed framework for modeling flood hazards, it presents significant limitations. Relying on a single, specific past flood event limits the ability of the method to fully capture the variability and complexity of flood dynamics across different seasons and/or meteorological conditions. The 2021 flood event may have been influenced by unique factors, such as specific rainfall patterns, soil moisture conditions, or river basin characteristics, which might not be representative of other potential flood scenarios.

By using only one event to build the catalog, the range of scenarios considered could be too narrow, failing to account for variations in flood magnitude, frequency, or duration that could arise from other types of extreme weather events. This may lead to a biased or incomplete hazard catalog, potentially underestimating the risk of future floods that differ from the 2021 event. Therefore, to enhance the robustness and reliability of flood hazard modeling in future work, we aim to incorporate multiple flood events with varying characteristics, rather than relying on a single past event. This would provide a more comprehensive and diverse range of flood scenarios for future hazard predictions.


\conclusions[Conclusions and Perspectives]  

This research presents a proof-of-concept for integrating Earth observation data into a Digital Twin framework dedicated to flood forecasting, demonstrated through the 2021 flood event in the Alzette catchment, G.D. of Luxembourg. By employing data assimilation techniques, particularly advanced particle filters, and leveraging Sentinel-1 satellite-derived flood probability maps, the system dynamically updates flood forecasts, enhancing their accuracy. The quantitative results show that assimilating remote sensing data improves flood forecasting in terms of water level and discharge predictions. This was also thanks to the satisfactory performance of the local LISFLOOD-FP hydraulic model that allows to generate a flood hazard datacube comprised of comprehensive range of flood scenarios at high spatial resolution.
Although some limitations exist, such as the under-prediction of flood peaks due to uncertainties in the GloFAS daily streamflow forecasts, the approach effectively reduces errors compared to open-loop forecasts. These uncertainties are an inherent trade-off for real-time forecasting, given the operational nature of the proposed Digital Twin framework for flood forecasting. 
The study demonstrates the potential of incorporating EO data and DA into real-time flood forecasting to improve disaster management and preparedness.

Future work should aim to refine the methodology by incorporating multiple past flood events into the flood hazard catalog to better capture a wider range of scenarios, improving forecast robustness. Additionally, updating the system to use the newer GloFAS 4.0 streamflow forecasts with higher spatial resolution could enhance forecast precision. Exploring alternative data sources and further enhancing the assimilation framework could also address some of the current limitations, such as the inability of Sentinel-1 to detect floodwater beneath dense vegetation, using for example L-band SAR satellites from new space missions like NiSAR (NASA \& ISRO) and ROSE-L (ESA).

Our next steps also undertake ‘what if’ scenarios to evaluate flood hazard and risk under different future climatic conditions and mitigation measures, using the same framework based on the flood hazard datacube pre-computed over the studied regions. Using near-future and mid-term climate change projections, the research assesses the potential impacts of changing precipitation and temperature patterns on flood dynamics. The development of a more comprehensive Digital Twin shall significantly improve flood resilience and early warning systems, contributing to better flood risk management.

\appendix
\section{GloFAS streamflow forecast initial conditions and forcing data}    
\label{app:glofas}

For the sake of completeness, this Appendix summarizes the initial conditions and forcing data used to generate the 51-member ensemble real-time daily streamflow forecasts for the next 30 days in GloFAS. Detailed information can be found on their website\footnote{\url{https://global-flood.emergency.copernicus.eu/technical-information/glofas-30day/}}.

\paragraph*{Hydro-meteorological initial conditions}
GloFAS real-time initial states (atmosphere and land surface) are updated daily, incorporating data from the atmosphere, land surface, and river states over the past five days, provided by the latest NRT GloFAS-ERA5 river discharge reanalysis \citep{harrigan2020glofas}.
However, to fill the gaps between the data from GloFAS-ERA5 and the real-time initialization of GloFAS forecasts, the control run 1-day forecasts from the ECMWF Integrated Forecast System (IFS) \citep{Balsamo2009}  from the preceding day are used as a fill-up, due to the latency in the availability of NRT reanalysis data \citep{harrigan2020daily}.

\paragraph*{ECMWF Numerical Weather Prediction (NWP) forcing}
GloFAS forecasts are generated using the most recent ensemble of NWP forecasts from the ECMWF IFS. 
The surface and sub-surface runoff within ECMWF IFS are generated using ECMWF's H-TESSEL land surface model, and then routed through the river network using the LISFLOOD hydrological model. 
The medium-range (5 to 10 days) and extended-range (10 to 30 days) ensemble runoff outputs are utilized for the 30-day GloFAS forecasts.

The ECMWF-ENS\footnote{\url{https://www.ecmwf.int/en/forecasts/documentation-and-support/medium-range-forecasts}} is ECMWF's ensemble forecasting system, comprising 51 members with a resolution of approximately 18 km (for day 1 to day 15), and doubling to 36 km (for day 16 to day 30). For GloFAS 30-day forecasts, outputs from the 00:00 UTC IFS medium-range runs are used daily up to 15 days, while the latest available IFS extended-range runs are used from day 16 to day 30. The 51 members (one control forecast and 50 forecasts) have slightly varied initial conditions and model physics. ECMWF-ENS is a probabilistic forecast system that shows the range of potential weather conditions up to 15 days in advance, including the likelihood of specific events like high winds or heavy rainfall. A larger spread among the ensemble members indicates higher uncertainty in the forecast.


\paragraph*{H-TESSEL (Hydrology-Tiled ECMWF Scheme for Surface Exchanges over Land)} 


H-TESSEL land surface model \citep{Balsamo2009} is an integral component of the ECMWF IFS. 
The land surface model provides atmospheric boundary conditions---such as heat, moisture, and momentum---by simulating the surface water and energy balances, along with the changes over time in soil temperature and moisture, snowpack, and vegetation interception.
Daily ensemble forecasts of surface and sub-surface runoff, including soil-to-groundwater percolation, are produced using the ECMWF-ENS outputs. These runoffs are then downscaled to a resolution of $0.1^\circ$ using the nearest neighbor method before being used as input for the LISFLOOD river routing model \citep{van2010lisflood}.

\paragraph*{LISFLOOD global}
LISFLOOD \citep{van2010lisflood} is a spatially distributed hydrological model based on GIS, featuring a 1D channel routing component. In GloFAS's global flood modeling framework, H-TESSEL handles the conversion of precipitation into surface and sub-surface runoff, while LISFLOOD is configured to simulate groundwater processes and flow routing. Surface runoff is transported via overland flow to the cell outlet, and sub-surface storage and movement are represented by two linear reservoirs. To improve the representation of hydrological variability, LISFLOOD's ground water and river routing parameters were calibrated using discharge time series from 1,287 catchments worldwide \citep{hirpa2018calibration}.

\clearpage
\section{Statistical values of discharge and water level at in-situ gauge station}     
\label{app2}
\begin{table}[h]
    \centering
    \caption{Return period statistical values (source: AGE).}
    \label{tab:return_period}
    \scalebox{0.9}{
    \begin{tabular}{c|cccc|cccc}
    \hline
    Discharge  [\Qunit] & Pfaffenthal & Schoenfels & Hunnebour & Bissen & Ettelbruck & Mersch & Steinsel & Walferdange \\
    & Alzette & Mamer & Eisch & Attert & Alzette & Alzette & Alzette & Alzette \\\hline
    HQ100 & 121 & 47.4 & 67.7 & 151 & 349 & 212 & 130 & - \\
    HQ50 & 107 & 41.6 & 58.8 & 131 & 311 & 189 & 116 & -\\
    HQ20 & 90.5 & 34.5 & 48.1 & 107 & 265 & 162 & 98.3 & -\\
    HQ10 & 78.5 & 29.8 & 40.6 & 90.3 & 232 & 142 & 85.5 & - \\
    HQ5 & 66.8 & 25.2 & 33.6 & 74.7 & 200 & 123 & 73.1 & -\\
    HQ2 & 51.8 & 19.5 & 25.2 & 55.9 & 158 & 96.8 & 56.9 & - \\\hline
    2003 observed peaks & 71.845 & 26.278 & 44.02 & 100.925 & 262.9 & 142.3 & \modif{82.7} & - \\\hline
    2021 observed peaks & 132.4 & 51.81 & 39.4 & 114.1  & 318.1 & 206.5 & 100.7 & - \\\hline 
\end{tabular}}

\vspace{1cm}

\scalebox{0.9}{
\begin{tabular}{c|cccc|cccc}
    \hline
    Water Level [cm] & Pfaffenthal &  Schoenfels & Hunnebour & Bissen  & Ettelbruck & Mersch & Steinsel & Walferdange \\
    & Alzette & Mamer & Eisch & Attert & Alzette & Alzette & Alzette & Alzette \\\hline
    HQ100 & 396 & 331 & 411 & 389 & 411 & 600 & 456 & -\\
    HQ50 & 372 & 314 & 391 & 365 & 384 & 574 & 445 & -\\
    HQ20 & 341 & 288 & 362 & 336 & 348 & 537 & 428 & -\\
    HQ10 & 307 & 260 & 335 & 314 & 317 & 498 & 411 & -\\
    HQ5 & 273 & 230 & 303 & 290 & 281 & 458 & 374 & -\\
    HQ2 & 227 & 189 & 252 & 258 & 232 & 400 & 322 & -\\\hline
    2003 observed peaks & 295 & 261 & 346 & 338 & 369 & 484 & 403 & 264\\\hline
    2021 observed peaks & 440 & 392 & 330 & 362 & 391 & 595 & 465 & 353 \\\hline
\end{tabular}}
\end{table}

\noappendix       




\appendixfigures  

\appendixtables   



\competinginterests{The authors declare that they have no conflict of interest.} 


\begin{acknowledgements}
This research work was funded by the LSA through an ESA Contract in the Luxembourg National Space programme (LuxIMPULSE), Contract No. 4000140965/23/NL/VR. The authors would like to acknowledge Dr. Antara Dasgupta for her contribution during her time at LIST. We would also thank Cyrille Tailliez (LIST) for his helps and supports with in-situ gauge data, and Dr. Tobias Stachl (EODC) and Dr. Yu Li (LIST) for their helps with GFM.
We would like to thank the Administration du Cadastre et de la Topographie (ACT, Luxembourg) for providing the 2019 LiDAR DTM over Luxembourg, and the Administration de la Gestion de l'Eau (AGE, Luxembourg) for providing the Alzette cross-section profiles within the "Cartographie des profils en travers 2021" program and for providing the meta-information on their in-situ gauge stations.
\end{acknowledgements}







\bibliographystyle{apalike}
\bibliography{./ref.bib}

\end{document}